\documentclass[final,3p]{elsarticle}

\usepackage{algorithm}
\usepackage{booktabs}
\usepackage{color}
\usepackage{amsthm}
\usepackage{amsmath,amssymb,amsfonts}
\usepackage{caption}
\usepackage{subcaption}

\newcommand{\beq}{\begin{equation}}
\newcommand{\eeq}{\end{equation}}

\usepackage{graphicx}
\usepackage{lineno,hyperref}
\modulolinenumbers[5]  

\newcommand{\ba}{\mbox{\boldmath $a$}}
\newcommand{\bb}{\mbox{\boldmath $b$}}

\newcommand{\bh}{\mbox{\boldmath $h$}}

\newcommand{\bu}{\mbox{\boldmath $u$}}
\newcommand{\bx}{\mbox{\boldmath $x$}}
\newcommand{\bw}{\mbox{\boldmath $w$}}
\newcommand{\bv}{\mbox{\boldmath $v$}}

\newcommand{\bA}{\mbox{\boldmath $A$}}
\newcommand{\bB}{\mbox{\boldmath $B$}}

\newcommand{\bD}{\mbox{\boldmath $D$}}
\newcommand{\bE}{\mbox{\boldmath $E$}}
\newcommand{\bG}{\mbox{\boldmath $G$}}
\newcommand{\bH}{\mbox{\boldmath $H$}}

\newcommand{\bN}{\mbox{\boldmath $N$}}
\newcommand{\bS}{\mbox{\boldmath $S$}}
\newcommand{\bU}{\mbox{\boldmath $U$}}
\newcommand{\bV}{\mbox{\boldmath $V$}}
\newcommand{\bW}{\mbox{\boldmath $W$}}
\newcommand{\bQ}{\mbox{\boldmath $Q$}}
\newcommand{\bX}{\mbox{\boldmath $X$}}
\newcommand{\bY}{\mbox{\boldmath $Y$}}
\newcommand{\bZ}{\mbox{\boldmath $Z$}}

\newcommand{\Real}{\mathbb R}

\newcommand{\be}{\begin{eqnarray}}
\newcommand{\ee}{\end{eqnarray}}

\newtheorem{definition}{Definition}
\newtheorem{remark}{Remark}

\biboptions{sort&compress}

\journal{arXiv}
\begin{document}

\begin{frontmatter}

\title{Non-negative tensor factorization for vibration-based local damage detection}

\author[label1]{Mateusz Gabor \corref{cor1}}
\cortext[cor1]{Corresponding author, mateusz.gabor@pwr.edu.pl}
\author[label1]{Rafal Zdunek}
\author[label2]{Radoslaw Zimroz}
\author[label2]{Jacek Wodecki}
\author[label3]{Agnieszka Wylomanska}

\address[label1]{Faculty of Electronics, Photonics, and Microsystems, Wroclaw University
of Science and Technology, Wroclaw, Poland.}

\address[label2]{Faculty of Geoengineering, Mining and Geology, Wroclaw University of Science and Technology, Na Grobli 15, 50-421 Wroclaw}

\address[label3]{Faculty of Pure and Applied Mathematics, Hugo Steinhaus Centre, Wroc{\l}aw University of Science and Technology, Janiszewskiego 14a, 50-370 Wroc{\l}aw, Poland}

\begin{abstract}
In this study, a novel non-negative tensor factorization (NTF)-based method for vibration-based local damage detection in rolling element bearings is proposed. As the diagnostic signal registered from a faulty machine is non-stationary, the time-frequency method is frequently used as a primary decomposition technique. It is proposed here to extract multi-linear NTF-based components from a 3D array of time-frequency representations of an observed signal partitioned into blocks. As a result, frequency and temporal informative components can be efficiently separated from non-informative ones. The experiments performed on synthetic and real signals demonstrate the high efficiency of the proposed method with respect to the already known non-negative matrix factorization approach.  
\end{abstract}


\end{frontmatter}

\section{Introduction}
\label{sec:introduciton}

Local damage detection in rotating machines is a challenge if the signal of interest (SOI) is hidden in a measured vibration signal. The SOI has specific properties (cyclic, impulsive, wide-band, weak energy, non-stationary, etc.), thus a time-frequency representation is often used as a transformation/decomposition technique. Till now, several effective approaches have been developed. As the signal is noisy, most of them before detection of SOI focus on signal enhancement by searching for specific patterns and extraction of them via various computational methods. Many approaches for determining informative frequency band (IFB) have been established (Spectral Kurtosis \cite{antoni2009cyclostationarity,antoni2006spectral,barszcz2009application}, Kurtogram \cite{antoni2007fast,Xiang2015}, Infogram \cite{antoni2016info} and its enhanced version \cite{hebda2022infogram}, SKRgram \cite{wang2016new}, spectral Gini index \cite{miao2017improvement_GINI}, IFBI$\alpha$-gram \cite{schmidt2020methodology}, spectral smoothness index \cite{bozchalooi2007smoothness}, harsogram \cite{zhao2016detection}, {IESFOgram \cite{mauricio2020improved}}, and many other statistical methods used for filter optimisation). A recent review can be found in \cite{hebda2020selection}.

Other powerful approach exploits cyclostationary properties of SOI \cite{antoni2004cyclostationary, antoni2009cyclostationarity, kruczek2020detect, KRUCZEK2021107737, WODECKI2021108400}. As a result, one obtains a bi-frequency map with information about modulating components and their carrier frequencies. More information on vibration damage detection is provided by Randall and Antoni in a very comprehensive tutorial on local damage detection in rolling element bearings \cite{randall2011rolling}.

As mentioned, the time-frequency methods are often used to analyze signal content in time and frequency domains. It can be the basis for estimating filter characteristics (above-mentioned selectors as spectral kurtosis) or the decomposition technique may be used as a filter directly (EMD, wavelets) \cite{lei2013review,peng2004application,feng2013recent}.
Many solutions have been developed for time-frequency based detection and decision-making, see for example \cite{FabienM}.
Recently, some novel ideas regarding time-frequency map analysis appeared in the literature. The spectrogram is a matrix of non-negative values describing energy at a given frequency and time instant. It can be assumed that some spectral bins are non-informative (do not contain SOI or signal-to-noise ratio (SNR) is too small to extract SOI effectively). Thus, a spectrogram has been considered as a matrix with the potential to reduce its dimensionality via principal component analysis (PCA) or other singular value decomposition-based feature extraction methods \cite{wodecki2016combination}. 

Non-negative matrix factorization (NMF) \cite{lee1999learning} is a powerful method for dimensionality reduction and feature extraction, which has been developed by many research groups for various applications \cite{8610086,9482220,8942862,Fu2019NonnegativeMF,Casalino2016}, including vibration signal processing techniques. 
Due to the parts-based property of the extracted features, NMF can separate mixed signals and extract intrinsic components that might carry meaningful information. Multiple models and algorithms for NMF \cite{cichocki2009nonnegative,Wang_2013,gillis2020nonnegative} allow us to factorize non-negative matrices in various ways and using various cost functions. 

Recently, some computational tools involving NMF have appeared to be very useful in condition monitoring. They have been applied to a spectrogram \cite{wodecki2019novel,Wodecki2020} as well as to the mentioned bi-frequency map \cite{wodecki2019impulsive}. As first applications of NMF to vibration signals analysis are very promising, we decided to explore deeper these opportunities. According to our findings, NMF-based procedure of spectrogram processing allows us to estimate a frequency range with informative content, so considering even only the frequency-related features, NMF can be used for selector estimation. Moreover, another NMF factor containing time-varying features allows us to find the temporal instants where the SOI appears in the observation. This is a way to extract information about SOI presence in the signal. If the components are separated properly, one could try to reconstruct partial spectrograms, or even to obtain a time series form of the individual components, e.g., using Griffin-Lim algorithm \cite{Wodecki2020,griffin1984signal}, under the minimal-phase condition assumption. 

One of the extensions of NMF is non-negative tensor factorization (NTF) \cite{Shashua2005,carroll1989fitting} that is addressed for multilinear feature extraction from non-negative data represented in the form of a multi-way array. This model has already found multiple applications in signal processing, see e.g., \cite{FitzGerald2008,6588559,8497054,app9183642}.
In this study, the spectrogram of a long one channel mixed vibration signal is sliced along the time axis into shorter segments, and the 3-way array is then processed with various NTF algorithms. In comparison to NMF, this multilinear approach allows us to better tailor the blind source separation tool for extracting the SOI from a mixed signal registered from rolling element bearings. We assume that the periodic and impulsive SOI is perturbed with a non-periodic sparse impulse noise or some non-stationary component, and a strong i.i.d. Gaussian noise. 
Cutting the spectrogram with the window whose length is adapted according to the information on the rotating speed and a design of a rolling element, the segments can be stacked to obtain a 3-way array. 
As a result, the frequency and temporal latent components of higher quality than in NMF can be extracted from multi-way observations, and additionally, this approach allows us to stronger relax the influence of the non-Gaussian perturbing signal that might have empirically different statistical properties in each slice. To extract the factor matrices in NTF, we used the $\beta$-divergence that is minimized with multiplicative update rules \cite{cichocki2009nonnegative}. The choice of this objective function is motivated by its higher flexibility in adapting to the separation problem with non-Gaussian residual errors. We demonstrated empirically the optimal cases for the analyzed synthetic and real vibration data. 

The paper is organized as follows. First, we will formulate the problem of local damage detection in the language of signal processing in Section \ref{sec:problem_form}. Section \ref{sec:nmf} presents the motivation behind using NMF for identifying the SOI in the observed one-channel mixed signal. The advantages of NTF over NMF and the algorithmic approach are studied in Section \ref{sec:tensor_background}. Next, the proposed NTF-based method for local damage detection is proposed in Section \ref{sec:method}. The next section describes the numerical experiments to obtain synthetic signals (simulations) and the experimental setup for vibration signal acquisition. Section \ref{sec:results} contains the outcome of the proposed method, both for simulations and real vibrations. The last subsection covers the discussion with a focus on the advantages of NTF. Finally, the conclusions are given in Section \ref{sec:conclusions}.  

\section{Problem formulation}
\label{sec:problem_form}
{\it Notation:} Multi-way arrays, matrices, vectors, and scalars are denoted by uppercase calligraphic letters (e.g., $\mathcal{X}$), uppercase boldface letters (e.g., $\bX$), lowercase boldface letters (e.g., $\bx$), and unbolded letters (e.g., $x$), respectively. Following the notation of Kolda \cite{Kolda08}, multi-way arrays will be referred to as tensors.

The vibration signal is frequently used to detect faults in mechanical systems. In some cases, due to early stage of damage (weak informative signal), complex design of the machine, or external, high amplitude disturbance, the informative component -- SOI, has significantly smaller amplitudes in comparison to the rest of the measured signal. 

\begin{figure}[h!]
    \centering
    \includegraphics[scale=0.15]{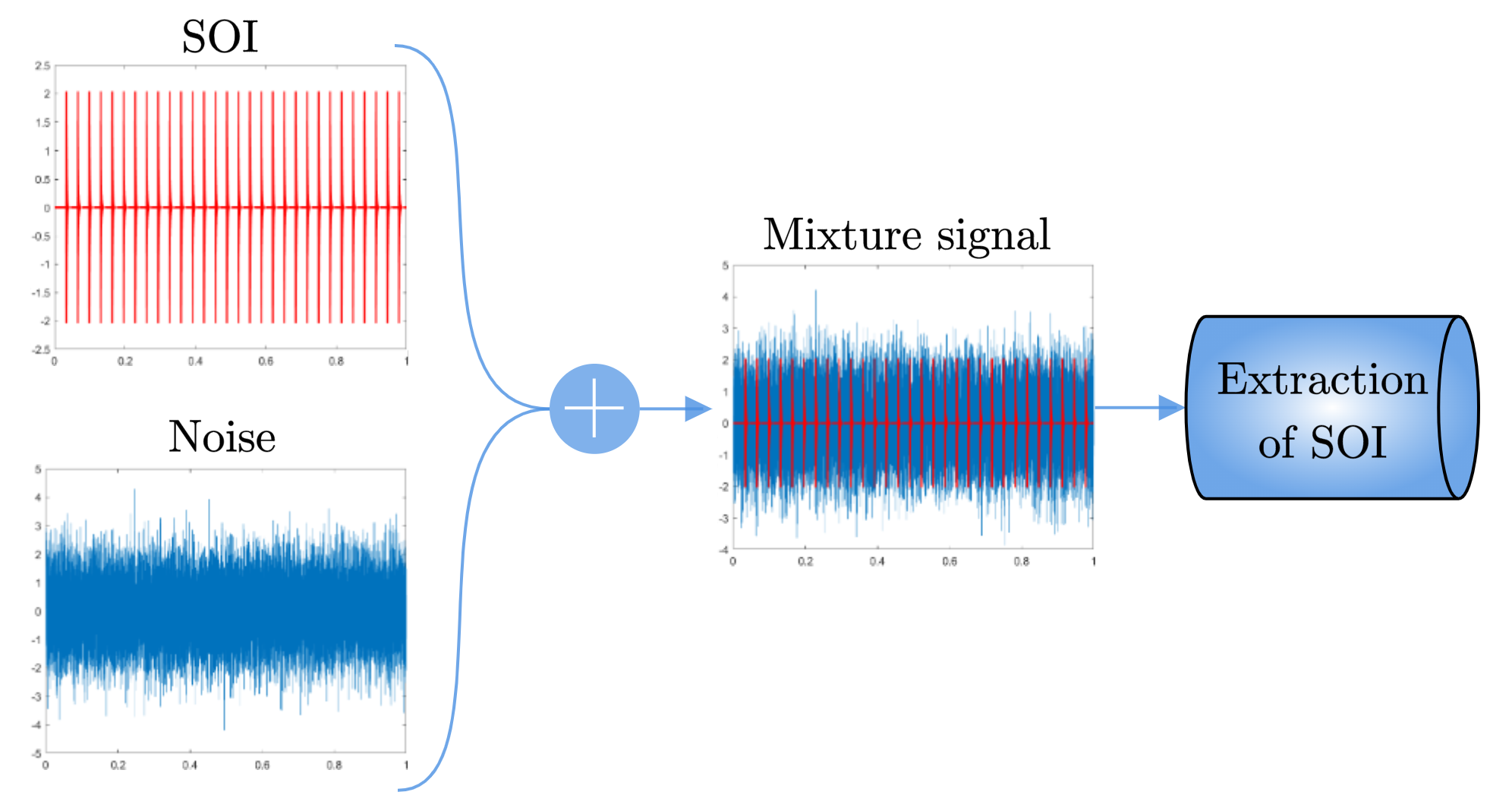}
    
    \caption{Graphical illustration of the problem: input data as an additive mixture of non-informative and informative components with poor SNR.}
    \label{fig:NTFx}
\end{figure}

We assume that observed signal $y(t)$ can be regarded as a superposition of three components: $s(t)$ -- periodic impulse SOI, $d(t)$ -- disturbing non-Gaussian (impulse) signal that might have empirically different statistical properties in each slice, and $n(t)$ -- disturbing zero-mean  Gaussian white noise.

Thus:
\be \label{eq_1} y(t) = s(t) + d(t) + n(t). \ee
Without loss of generality, let us assume that signal $s(t)$ can be modelled by a damped sinusoidal signal repeated with period $T_p$. Hence: 
\be \label{eq_2} s(t) = S_a\sum_{m = -\infty}^{\infty} \exp \{ \alpha_s t\} \sin \left (2 \pi f_c t + \phi_s \right ) \ast \delta(t - mT_p), \ee
where $S_a$ is the magnitude of the sinusoidal signal, $f_c$ is its frequency, $\alpha_c$ is the damping factor, $\delta(\cdot)$ is the Dirac delta function, and symbol $\ast$ is the convolutional operator. 

Applying the short-time Fourier transform (STFT) to (\ref{eq_2}) under the assumption that $\frac{1}{2f_c} < M << T_p$, where $M$ is the window length, we get the time-frequency representation whose squared magnitude $\mathcal{S}(f,\tau)$ (spectrogram) takes the form: 
\be \label{eq_3} \mathcal{S}(f,\tau) & = & |STFT \{ s(t)\}(f,\tau) |^2 \nonumber \\ & = &  \sum_{m = -\infty}^{\infty} \frac{S_a^2 \psi(\tau)}{8 \pi \left ( \alpha_s^2 + 4 \pi^2 (f - f_c)^2 \right )} \ast \delta \left (\tau - \frac{m T_p}{M} \right ), \ee 
where $\psi(\tau) > 0$ is the dispersion function determined by the window function in the STFT. Spectrogram $\mathcal{S}(f,\tau)$ can be regarded as a temporal comb of unimodal spectral function 
$w(f,\tau) = \frac{S_a^2 \psi(\tau)}{8 \pi \left ( \alpha_s^2 + 4 \pi^2 (f - f_c)^2 \right )}$ damped by factor $\psi(\tau)$ with respect to the time. 

Let $\mathcal{Y}(f,\tau)$, $\mathcal{D}(f,\tau)$, and $\mathcal{N}(f,\tau)$ be the spectrograms of $y(t)$, $d(t)$, and $n(t)$, respectively. Due to the linearity property of STFT, we have the superposition of spectrograms: 
\be \label{eq_4} \mathcal{Y}(f,\tau) = \mathcal{S}(f,\tau) + \mathcal{D}(f,\tau) + \mathcal{N}(f,\tau). \ee
The aim is to get the relevant information on existence of $\mathcal{S}(f,\tau)$ in $\mathcal{Y}(f,\tau)$.

\section{Non-negative matrix factorization-based analysis}
\label{sec:nmf}

It is easy to notice that $\mathcal{S}(f,\tau)$ in (\ref{eq_3}) can be presented in the factorized form: 
\be \label{eq_5} \mathcal{S}(f,\tau) = W(f)H(\tau), \ee
where $W(f) = \frac{S_a^2}{8 \pi \left ( \alpha_s^2 + 4 \pi^2 (f - f_c)^2 \right )} > 0$ is a positive-value function that represents the spectral component, and $H(\tau) = \sum_{m = -\infty}^{\infty} \psi(\tau) \ast \delta \left (\tau - \frac{m T_p}{M} \right ) \geq 0$ is a non-negative-value function that expresses the temporal comb component.  

In the discrete representation, $W(f)$ takes the form of vector $\bw_c = [w_i^{(c)}] \in \Real_+^I$, where $w_i^{(c)} = \frac{S_a^2}{8 \pi \left ( \alpha_s^2 + 4 \pi^2 (\Delta f i - f_c)^2 \right )}$,  $i = 1, \ldots, I$, and $\Delta f$ is the frequency step. Similarly, the temporal component in the discrete time representation may be given by: 
$\bh_c = [h_k^{(c)}] \in \Real_+^{K}$, where 
\be \label{eq5a} h_k^{(c)} = \sum_{m = -\infty}^{\infty} \psi \left (\frac{k}{f_s} \right ) \ast \delta_c \left (\frac{k}{f_s} - \frac{m T_p}{Mf_s} \right ), \ee where $\delta_c(\cdot)$ is the Kronecker delta function, and $f_s$ is the sampling frequency.   

 Let $\bS = [s_{ik}] \in \Real_+^{I \times K}$ be the discrete version of $\mathcal{S}(f,\tau)$. It can therefore be expressed by the rank-one matrix approximation model: 
$\bS = \bw_c \bh_c^T \in \Real^{I \times K}$. 
Assuming spectrogram $\mathcal{D}(f,\tau)$ in its discrete representation has the similar form:  $\bD = \bw_d \bh_d^T \in \Real^{I \times K}$, and $\bN$ be a discrete version of $\mathcal{N}(f,\tau)$ that can be approximated by a superposition of $J$ rank-one components, i.e.,  $\bN \cong \sum_{j = 1}^{J_n} \bw_j \bh_j^T$ due to more complex nature of a Gaussian noise. Hence, considering model (\ref{eq_4}), the discrete version of the observed spectrogram can be expressed by model: 
\be \label{eq_6} \bY \cong \sum_{j = 1}^J \bw_j \bh_j^T, \ee where $J = J_n + 2$ and $\bw_{J_n + 1} = \bw_d$, $\bw_{J_n + 2} = \bw_c$, $\bh_{J_n + 1} = \bh_d$, and $\bh_{J_n + 2} = \bw_c$. Model (\ref{eq_6}) can be equivalently expressed by the following: 
\be \label{eq_7} \bY \cong \bW \bH^T, \ee where 
$\bW \in \Real_+^{I \times J}$ is a non-negative matrix representing frequency components and $\bH \in \Real_+^{K \times J}$ is a non-negative matrix containing temporal components. 

Note that $\bY$ in (\ref{eq_7}) has the form of the fundamental NMF model, which motivates the usage of NMF for solving the main problem in this study. The concept of using NMF for vibration-based local damage detection in rolling element bearings has already been analyzed in the following papers \cite{wodecki2019novel,Wodecki2020,wodecki2019impulsive}.

In this study, we extend this concept by using a multilinear approach to detect the cyclic impulsive signal, i.e., signal $s(t)$. The standard model of NMF assumes that spectral component $\bw_c$ exists directly in the spectrogram of the observed mixed signal across the temporal axis or can be extracted as a predominant component in a conic combination of other spectral components. Geometrically, this problem can be regarded as finding the extreme ray in the convex cone of observation points. In this approach, each observed sample contributes equally to the process of finding the extreme rays, which may lead to strong influence of non-periodic impulse perturbation -- signal $d(t)$ in (\ref{eq_1}) -- onto the desired signal $s(t)$. To relax this problem, we may divide the whole observed mixed signal $y(t)$ into segments, and to assign different weights to each segment. This concept  results from the assumption that $d(t)$ in (\ref{eq_1}) is a non-stationary transient signal or a non-period impulse signal that may have different empirical properties in each segment, and hence it may be better modeled in each segment than in the whole observation window. This approach motivates the usage of the NTF-based feature extraction method to the primary task in this study. 

\section{Non-negative tensor factorization-based analysis}
\label{sec:tensor_background}

Tensors can be interpreted as multi-way arrays that generalize  vectors and matrices to higher dimensions. In the language of mathematics, let $\mathcal{X} = [x_{i_1,\ldots,i_N}] \in \Real^{I_1 \times \dots \times I_N}$ be the $N$-order tensor, where $I_n$ denotes the number of entries across the $n$-th mode. For example, vectors and matrices are the 1-order and the 2-order tensors, respectively. 

\begin{definition} \label{def_1}
Mode-$n$ unfolding of $N$-way array ($N$-modal tensor) $\mathcal{X} \in \Real^{I_1 \times \ldots \times I_N}$ reshapes it to matrix $\bX_{(n)} = [x_{i_n,j}] \in \Real^{I_n \times \prod_{p \not = n} I_p}$ for $n \in \{1, \ldots, N \}$, where
$j = 1 + \sum_{\substack{k=1 \\ k \neq n}}^N(i_k-1)j_k$ with  $j_k = \prod_{\substack{m=1 \\ m \neq n}}^{k-1}I_m$ and $i_n = 1, \ldots, I_n$. 
\end{definition}

\begin{definition} \label{def_2}
Mode-$n$ product (also known as the tensor-matrix product): Let $\mathcal{X} \in \Real^{I_1 \times \ldots \times I_N}$ and $\bU \in \Real^{J \times I_n}$. Then the mode-$n$ product of tensor $\mathcal{X}$ with matrix $\bU$ is defined by $\mathcal{Z} =  \mathcal{X} \times_n \bU$, where $$\mathcal{Z} = [z_{i_1,\ldots,i_{n-1},j,i_{n+1},\ldots,i_N}] \in \Real^{I_1 \times \ldots \times I_{n-1} \times J \times I_{n+1} \times \ldots \times I_N}$$ and
$$z_{i_1,\ldots,i_{n-1},j,i_{n+1},\ldots,i_N} = \sum_{i_n=1}^{I_n} x_{i_1,i_2,\ldots,i_N} u_{j,i_n}.$$
\end{definition}

\begin{definition} \label{def_3}
Let $\bA = [\ba_1, \ldots, \ba_J] \in \Real^{P \times J}$ and $\bB = [\bb_1, \ldots, \bb_J] \in \Real^{R \times J}$. The Khatri-Rao product of $\bA$ and $\bB$ is defined by:
\be \label{eq_ten_1} \bA \odot \bB = \left [\ba_1 \otimes \bb_1, \ldots, \ba_J \otimes \bb_J \right ] \in \Real^{PR \times J}, \ee
where $\otimes$ denotes the Kronecker product. 
\end{definition}

\subsection{Non-negative tensor factorization}
\label{sec:ntf}

\begin{figure}[h!]
    \centering
    \includegraphics[scale=0.25]{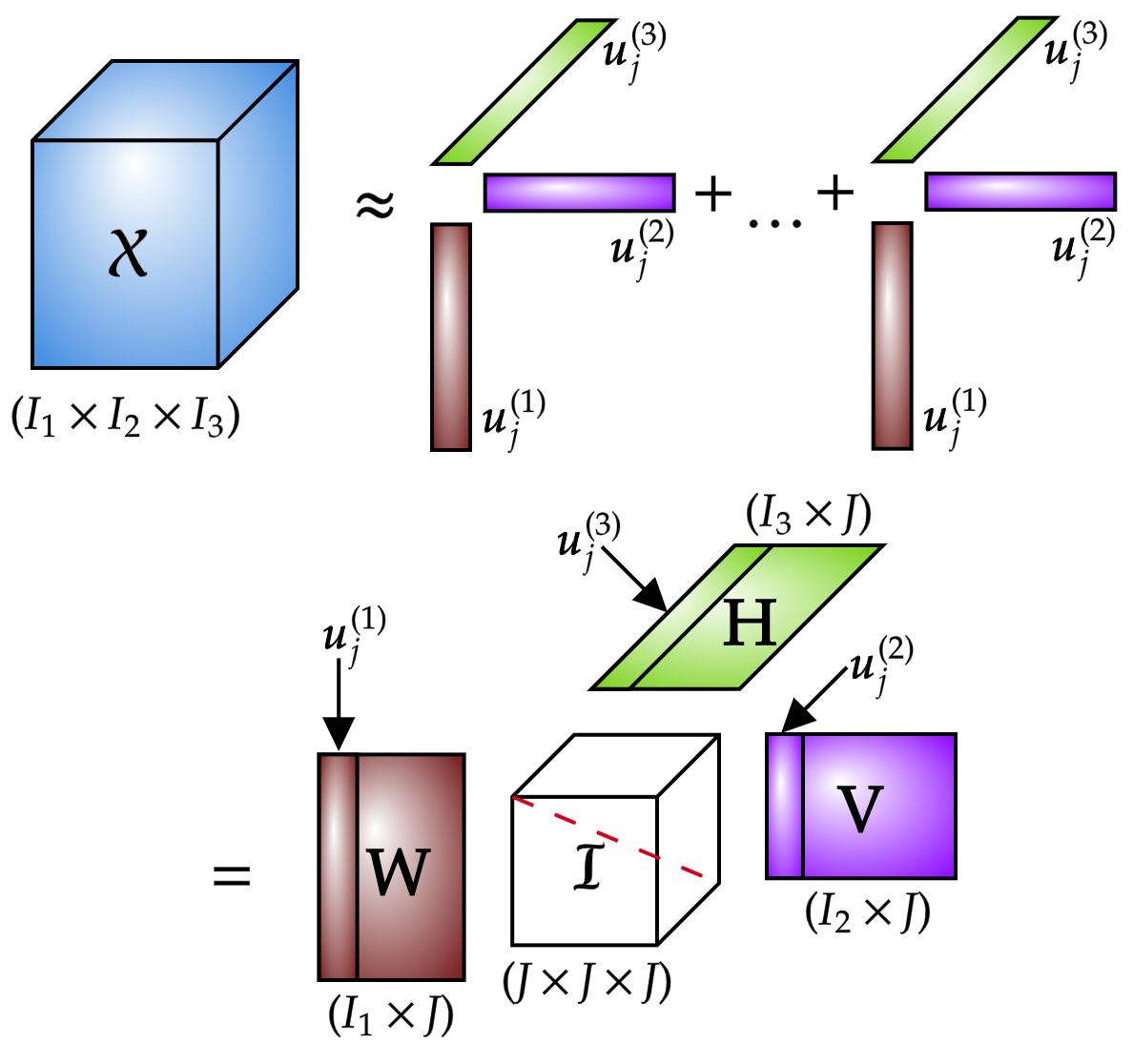}
    \caption{NTF factorization of the 3-order tensor.}
    \label{fig:NTF}
\end{figure}

Non-negaive tensor factorization (NTF) \cite{Shashua2005} is a multi-way extension of NMF for representing non-negatively constrained multi-way data with a low-rank latent structure. It can also be regarded as a particular case of the CANDECOMP/PARAFAC (CP) model \cite{Carroll_Chang_70,Harshman1970}, where the non-negativity constraints are imposed onto all the estimated factors. NTF of multi-way data given in the form of  
$\mathcal{X} \in \Real_+^{I_1 \times \ldots \times I_N}$ can be expressed by the model: 
\begin{equation}
\label{eq_ntf_1}
    \mathcal{X} = \sum_{j=1}^J \bu_j^{(1)}\circ \bu_j^{(2)} \circ \cdots \circ \bu_j^{(N)},
\end{equation}
where $\bu_j^{(n)} \in \Real_+^{I_n}$ is the $j$-th latent component representing the non-negative features extracted from $\mathcal{X}$ across its $n$-th mode ($n = 1, \ldots, N$), $J$ denotes the rank of factorization, and $\circ$ means the outer product. Model (\ref{eq_ntf_1}) can be equivalently rewritten in a more compact form: \begin{equation}
\label{eq_ntf_2}
    \mathcal{X} = \mathcal{I} \times_1 \bU^{(1)} \times_2 \bU^{(2)} \times_3 \cdots \times_N \bU^{(N)} ,
\end{equation}
where $\bU^{(n)} = [\bu_1^{(n)}, \ldots, \bu_J^{(n)}] \in \Real_+^{I_n \times J}$ is the factor matrix containing non-negative features captured across the $n$-th mode, $\mathcal{I} \in \Real^{J \times \ldots \times J}$ is the superdiagonal identity tensor, and $\times_n$ stands for the mode-$n$ product given in Def. \ref{def_2}. The NTF model is illustrated in Fig. \ref{fig:NTF}.

Applying the mode-$n$ unfolding to (\ref{eq_ntf_2}) according to Def. \ref{def_1}, we get:
\be \label{eq_ntf_unfold} \forall n: \bX_{(n)} & = & \bU^{(n)} \left (\bU^{(N)} \odot \cdots \odot \bU^{(n+1)} \odot \bU^{(n-1)} \odot \cdots \odot \bU^{(1)} \right )^T \nonumber \\
& = & \bU^{(n)} \bB^{(n)} \in \Real_+^{I_n \times \prod_{p \not = n} I_p}, \ee
where $\bB^{(n)} = \left (\bU^{(N)} \odot \cdots \odot \bU^{(n+1)} \odot \bU^{(n-1)} \odot \cdots \odot \bU^{(1)} \right )^T$. 

\subsection{Tensorization of observations}
\label{sec:tensorization}

Let us assume that observed spectrogram $\bY$ in (\ref{eq_7}) can be partitioned along the time axis into $L$ blocks of $P$ samples, i.e., 
\be \label{eq_ntf_unfold_Y} \bY  = \left [\bY_1, \bY_2, \ldots, \bY_L \right ] \in \Real_+^{I \times K}, \ee where $\bY_l \in \Real_+^{I \times P}$ is a lateral slice for $l = 1, \ldots, L$ and $K = LP$. Matrix $\bY$ can be regarded as the mode-1 unfolding of the 3-way tensor $\mathcal{Y} \in \Real_+^{I \times P \times L}$, i.e., $\bY \equiv \bY_{(1)}$.     

Vector $\bh_c$ according to (\ref{eq5a}) represents a comb of pulses repeated with period $\frac{T_p}{Mf_s}$. Knowing $T_p$, the length of each block can be selected as $M_p$ multiplicity of $\frac{T_p}{Mf_s}$. Thus $P = \frac{M_pT_p}{Mf_s}$, and hence $h_k^{(c)}$ in each block takes the form:
\be \label{eq5a_ntf} h_p^{(c)} = \sum_{m = 0}^{M_p} \psi \left (\frac{p}{f_s} \right ) \ast \delta_c \left (\frac{p}{f_s} - \frac{m T_p}{Mf_s} \right ) \ee for $p = 1, \ldots, P$. 
Considering (\ref{eq_6}), model (\ref{eq_ntf_unfold_Y}) can be rewritten as \be \label{eq_ntf_unfold_Yn} 
\bY  = \left [\bw_c \tilde{\bh}_c^T + \tilde{\bY}_1, \bw_c \tilde{\bh}_c^T + \tilde{\bY}_2, \ldots, \bw_c \tilde{\bh}_c^T + \tilde{\bY}_L \right ] \in \Real_+^{I \times K}, \ee where $\tilde{\bh}_c = [h_p^{(c)}] \in \Real_+^{P}$, $\tilde{\bY}_l \cong \sum_{j = 1}^{J_n+1} v_{lj} \bw_j \tilde{\bh}_j^T$ represents the total disturbances in the $l$-th lateral slice of $\mathcal{Y}$, and $v_{lj} > 0$ is the scaling factor that is a data driven parameter estimated individually for each block and each component. Tensorizing model (\ref{eq_ntf_unfold_Yn}) by using the inverse operation to the mode-1 unfolding, we get the following model:
\be \label{eq_ntf_main} \mathcal{Y} \cong \sum_{j = 1}^{J_n} \bw_j \circ \tilde{\bh}_j \circ \bv_j + \bw_d \circ \tilde{\bh}_d \circ \bv_d  + \bw_c \circ \tilde{\bh}_c \circ \bv_c, \ee
where $\bv_d = [v_1^{(d)}, \ldots, v_L^{(d)}] \in \Real_+^{L}$ and $\bv_c = [v_1^{(c)}, \ldots, v_L^{(c)}] \in \Real_+^{L}$ are the vectors of scaling factors for the disturbing non-periodic impulse signal and the desired periodic impulse signal, respectively. If signal $d(t)$ does not occur in the $l$-th block, then $v_l^{(d)} \rightarrow 0$. A long-period stationarity of $s(t)$ should lead to $v_l^{(c)} \rightarrow 1$ for each $l$. Creating matrices $\bW = [\bw_1, \ldots, \bw_{J_n}, \bw_d, \bw_c] \in \Real_+^{I \times J}$, $\tilde{\bH} = [\tilde{\bh}_1, \ldots, \tilde{\bh}_{J_n}, \tilde{\bh}_d, \tilde{\bh}_c] \in \Real_+^{P \times J}$, and $\bV = [\bv_1, \ldots, \bv_{J_n}, \bv_d, \bv_c] \in \Real_+^{L \times J}$, formula (\ref{eq_ntf_main}) can be rewritten to the following model: 
\be \label{eq_ntf_main_mtx} \mathcal{Y} \cong \mathcal{I} \times_1 \bW \times_2 \tilde{\bH} \times_3 \bV. \ee
Note that model (\ref{eq_ntf_main_mtx}) has the same form as the generic NTF model in (\ref{eq_ntf_2}). 

\subsection{Algorithmic approach}
\label{sec:algorithm}

The factors in model (\ref{eq_ntf_main_mtx}) can be estimated using various objective functions, resulting from the assumption on a distribution of observations. The $\beta$-divergence covers a wide family of well-known dissimilarity measures used as objective functions in various tensor decomposition methods, including the CP model. For model (\ref{eq_ntf_main_mtx}), it can be expressed by the following form:
\be \label{eq_beta_div}
\Psi^{(beta)} = \left \{ \begin{array}{cc} 
\displaystyle{ \sum_{i,p,l} \left (y_{ipl} \frac{y_{ipl}^{\beta} - q_{ipl}^{\beta}}{\beta} - \frac{y_{ipl}^{\beta+1} - q_{ipl}^{\beta+1}}{\beta+1}   \right )} & \beta  > 0, \\
\displaystyle{ \sum_{i,p,l} \left (y_{ipl} \ln \left(\frac{y_{ipl}}{q_{ipl}} \right) - y_{ipl} + q_{ipl}   \right )} & \beta = 0, \\
\displaystyle{ \sum_{i,p,l} \left ( \ln\left(\frac{q_{ipl}}{y_{ipl}} \right)  + \left(\frac{y_{ipl}}{q_{ipl}} \right) - 1 \right )} & \beta = -1,
\end{array}  \right .
\ee
where $\mathcal{Y} = [y_{ipl}]$, $\mathcal{Q} = [q_{ipl}]$,  and $q_{ipl} = \sum_{j = 1}^J w_{ij} \tilde{h}_{pj} v_{lj}$ is the model given in (\ref{eq_ntf_main_mtx}). When $\beta = 0$, the $\beta$-divergence in (\ref{eq_beta_div}) takes the form of the generalized Kullback-Leibler (KL) divergence that is optimal for the Poisson-distributed data.  If $\beta = -1$, the $\beta$-divergence becomes the Itakura-Saito (IS) distance that is the most suitable for the Gamma noise, and the standard Euclidean distance can be obtained for $\beta = 1$.  

Function (\ref{eq_beta_div}) can be minimized with various algorithmic approaches; almost all of them are based on the gradient-based updates. Examples include the multiplicative or quasi-Newton update rules that can be regarded as scaled gradient descent rules. The gradients of (\ref{eq_beta_div}) with respect to $\bW$, $\tilde{\bH}$, and $\bV$ have the forms: 
\be \label{eq_grad_W_ntf}
\bG_W^{(NTF)} & = & \nabla_{\bW}^2 \Psi^{(beta)} = \left(\bQ_{(1)}^{.\beta} - \bY_{(1)}  \circledast \bQ_{(1)}^{.\beta -1}\right) \left (\bV \odot \tilde{\bH} \right ) \nonumber \\
& = & \left ( \bW(\bV \odot \tilde{\bH})^T \right )^{.\beta}\left (\bV \odot \tilde{\bH} \right )   - \bZ_{(1)} \left (\bV \odot \tilde{\bH} \right ), 
\ee
\be \label{eq_grad_H_ntf}
\bG_H^{(NTF)}  =  \left ( \tilde{\bH}(\bV \odot \bW)^T \right )^{.\beta}\left (\bV \odot \bW \right )  - \bZ_{(2)} \left (\bV \odot \bW \right ), \ee
\be \label{eq_grad_V_ntf}
\bG_V^{(NTF)}  = \left ( \bV(\tilde{\bH} \odot \bW)^T \right )^{.\beta}\left (\tilde{\bH} \odot \bW \right )  - \bZ_{(3)} \left (\tilde{\bH} \odot \bW \right ), \ee
where $\mathcal{Q} = [q_{ipl}] \in \Real^{I \times P \times L}$, $\mathcal{Z} = \mathcal{Y} \circledast  \mathcal{Q}^{.\beta - 1}$, $\circledast$ stands for the Hadamard product, and $(\cdot)^{.\beta}$ denotes the element-wise power to $\beta$.  

All the gradients can be expressed in the form of a subtraction of two components, e.g., $\bG_W^{(NTF)} = \bG_W^{(1)} - \bG_W^{(2)}$ for  (\ref{eq_grad_W_ntf}), where 
\be \label{eq_grad_W_ntf_1}
\bG_W^{(1)} = \left ( \bW(\bV \odot \tilde{\bH})^T \right )^{.\beta}\left (\bV \odot \tilde{\bH} \right ), \ee
and $\bG_W^{(2)} = \bZ_{(1)} \left (\bV \odot \tilde{\bH} \right )$. 

In the experiments, we used the multiplicative update rules derived from the standard gradient descent rule with a suitably chosen step length. As a result, the multiplicative update rules for minimization of objective function (\ref{eq_beta_div}) with respect to $\bW$, $\tilde{\bH}$, and $\bV$ take the forms:
\be \label{eq_multi_W}
\bW & \leftarrow & \bW \circledast \bG_W^{(2)} \oslash \bG_W^{(1)}, \\
\label{eq_multi_H}
\tilde{\bH} & \leftarrow & \tilde{\bH} \circledast \bG_H^{(2)} \oslash \bG_H^{(1)}, \\
\bV & \leftarrow & \bV \circledast \bG_V^{(2)} \oslash \bG_V^{(1)}, \label{eq_multi_V}
\ee
where symbol $\oslash$ denotes the element-wise division, and matrices $\bG_H^{(1)}$ and $\bG_H^{(2)}$ come from (\ref{eq_grad_H_ntf}), and $\bG_V^{(1)}$ and $\bG_V^{(2)}$ result from (\ref{eq_grad_V_ntf}). 

\begin{remark} \label{r0}
Note that $\frac{\partial}{\partial q_{ipl}} \Psi^{(\beta = 1)} = q_{ipl} - y_{ipl}$, $\frac{\partial}{\partial q_{ipl}} \Psi^{(\beta = 0)} = \frac{q_{ipl} - y_{ipl}}{q_{ipl}}$, and $\frac{\partial}{\partial q_{ipl}} \Psi^{(\beta = -1)} = \frac{q_{ipl} - y_{ipl}}{q_{ipl}^2}$. Decreasing $\beta$, the update rules in (\ref{eq_multi_W})--(\ref{eq_multi_V}) become less sensitive to high-energy components due to the gradient scaling. In particular, the Itakura-Saito distance ($\beta = -1$) is scale-invariant and particularly efficient for recovering low-energy components, usually located in a higher frequency band. This is the case in the discussed application, where the SOI has a lower energy than the noisy perturbations and is usually located in a higher frequency band. 
\end{remark}

\subsection{NTF vs. NMF}
\label{sec:ntf_nmf}

The advantages of using NTF for the primary task and the motivation behind it can be demonstrated by analyzing the gradient behavior of the update rules of NTF versus NMF. 
For simplicity reasons, we analyze the case of $\beta = 1$, i.e., when the $\beta$-divergence simplifies to the Euclidean (EU) distance. Consequently, component $\bG_W^{(1)}$ in (\ref{eq_grad_W_ntf_1}) takes the form:
\be \label{eq_grad_W_ntf_2}
\bG_W^{(1)} = \bW \left(\bV^T\bV \circledast \tilde{\bH}^T\tilde{\bH} \right ) = \bW \left (\bS_v  \circledast \tilde{\bH}^T\tilde{\bH} \right ), \ee
where $\bS_v = \bV^T\bV \in \Real_+^{J \times J}$ is the weight matrix that performs element-wise weighting. 
Considering the fundamental properties of the Khatri-Rao product, $\bG_W^{(2)}$ in (\ref{eq_grad_W_ntf}) can be rewritten to the form: \be \label{eq_ntf_nmf_2} \bG_W^{(2)} & = & \bY_{(1)} (\bV \odot \tilde{\bH})  =  \bY_{(1)} \left [\tilde{\bH} {\rm diag} \{ \bv_1 \}; \ldots; \tilde{\bH} {\rm diag} \{ \bv_L \} \right ] \nonumber \\
& = & \sum_{l = 1}^L \bY_l \left(\tilde{\bH} {\rm diag} \{ \bv_l \}\right)  =   \sum_{l = 1}^L \bY_l \bar{\bH}_l \in \Real_+^{I \times J}, 
\ee
where $\bY_l \in \Real_+^{I \times P}$ is the $l$-th lateral slice of $\mathcal{Y}$, $\bv_l \in \Real^{1 \times J}$ is the $l$-th row vector of $\bV$, and $\bar{\bH}_l$ is column-scaled  matrix $\tilde{\bH}$ with vector $\bv_l$. Following the similar operations, gradients (\ref{eq_grad_H_ntf}) and (\ref{eq_grad_V_ntf}) simplify to the forms:
\be \label{eq_grad_H_ntf_eu}
\bG_H^{(NTF)}  =  \tilde{\bH} \left (\bS_v  \circledast \bW^T \bW \right ) - \sum_{l = 1}^L \bY_l^T \bar{\bW}_l, \ee
\be \label{eq_grad_V_ntf_eu}
\bG_V^{(NTF)}  =  \bV \left (\tilde{\bH}^T\tilde{\bH}  \circledast \bW^T \bW \right ) - \bY_{(3)}(\tilde{\bH} \odot \bW), \ee
where $\bar{\bW}_l$ is column-scaled matrix $\bW$ with vector $\bv_l$. 

Applying the NMF model to (\ref{eq_ntf_unfold_Y}), we have:
\be \label{eq_nmf_Y} \bY  = \left [\bY_1, \ldots, \bY_L \right ] = \bW \left [\bH_1^T, \ldots, \bH_L^T \right ], \ee
where $\bH_l \in \Real_+^{P \times J}$ is the $l$-th block of $\bH$ corresponding to block $\bY_l$. Following a similar strategy with respect to model (\ref{eq_nmf_Y}), the gradients of the Euclidean distance with respect to matrices $\bW$ and $\bH_l$ can be formulated as follows:
\be \label{eq_grad_W_nmf}
\bG_W^{(NMF)} & = &  \left(\bW\bH^T - \bY \right) \bH = \bW \sum_{l = 1}^L \bH_l^T \bH_l - \sum_{l = 1}^L \bY_l \bH_l.
\ee
\be \label{eq_grad_H_nmf}
\bG_H^{(NMF)} & = &  \left(\bH_l\bW^T - \bY_l^T \right)\bW = 
\bH_l\bW^T\bW - \bY_l^T \bW.
\ee

Assuming $\tilde{\bH} = \sum_{l = 1}^L \bH_l$, it is easy to notice that gradients $\bG_W^{(NTF)}$ and $\bG_W^{(NMF)}$ differ only by the weighting factor $\bS_v$ and the scaling of factor $\bW$. Similarly, gradient (\ref{eq_grad_H_ntf_eu}) is an extended version of (\ref{eq_grad_H_nmf}) by considering weighting factor $\bS_v$, the scaling of $\bH_l$, and the superposition over $L$ slices. If $\bV = \bE_{L \times J}$, i.e., a matrix of all ones, $\bS_v = \bE_{J \times J}$, there is no difference between the update rules for NTF and NMF. This means that NMF can be considered a special case of NTF when $\bV = \bE_{L \times J}$ and $\tilde{\bH} = \sum_{l = 1}^L \bH_l$. 

\begin{remark} \label{r1}
Since matrix $\bV$ in (\ref{eq_ntf_main_mtx}) is not a matrix of all ones but computed with update rule (\ref{eq_multi_V}) that adapts the rows in $\bV$ with the lateral slices in $\mathcal{Y}$, NTF cannot be considered here as an equivalent model to NMF. Furthermore, due to the mentioned weighting factor $\bS_v$ and the special scaling, NTF is more flexible and adapts better to non-stationary and transient components in $\mathcal{Y}$. For example, if signal $d(t)$ is not observed in the $l$-th slice of $\mathcal{Y}$, then the $l$-th entry of vector $\bv_d$ in (\ref{eq_ntf_main})  should go to zero, and this component should not affect the updates for $\bW$ and $\tilde{\bH}$ according to rules (\ref{eq_multi_W}) and (\ref{eq_multi_H}). This is an important advantage of NTF over NMF, which motivated our study.
\end{remark}

\section{Fault detection in rolling element bearings using NMF and NTF}
\label{sec:method}
The idea of using NMF and NTF for fault detection is based on the blind source separation approach, in which decomposition rank $J$ of NMF and NTF determines how many source components (classes) are to be separated from the observed mixed signal.

\begin{figure}[h!]
    \centering
    \includegraphics[scale=0.135]{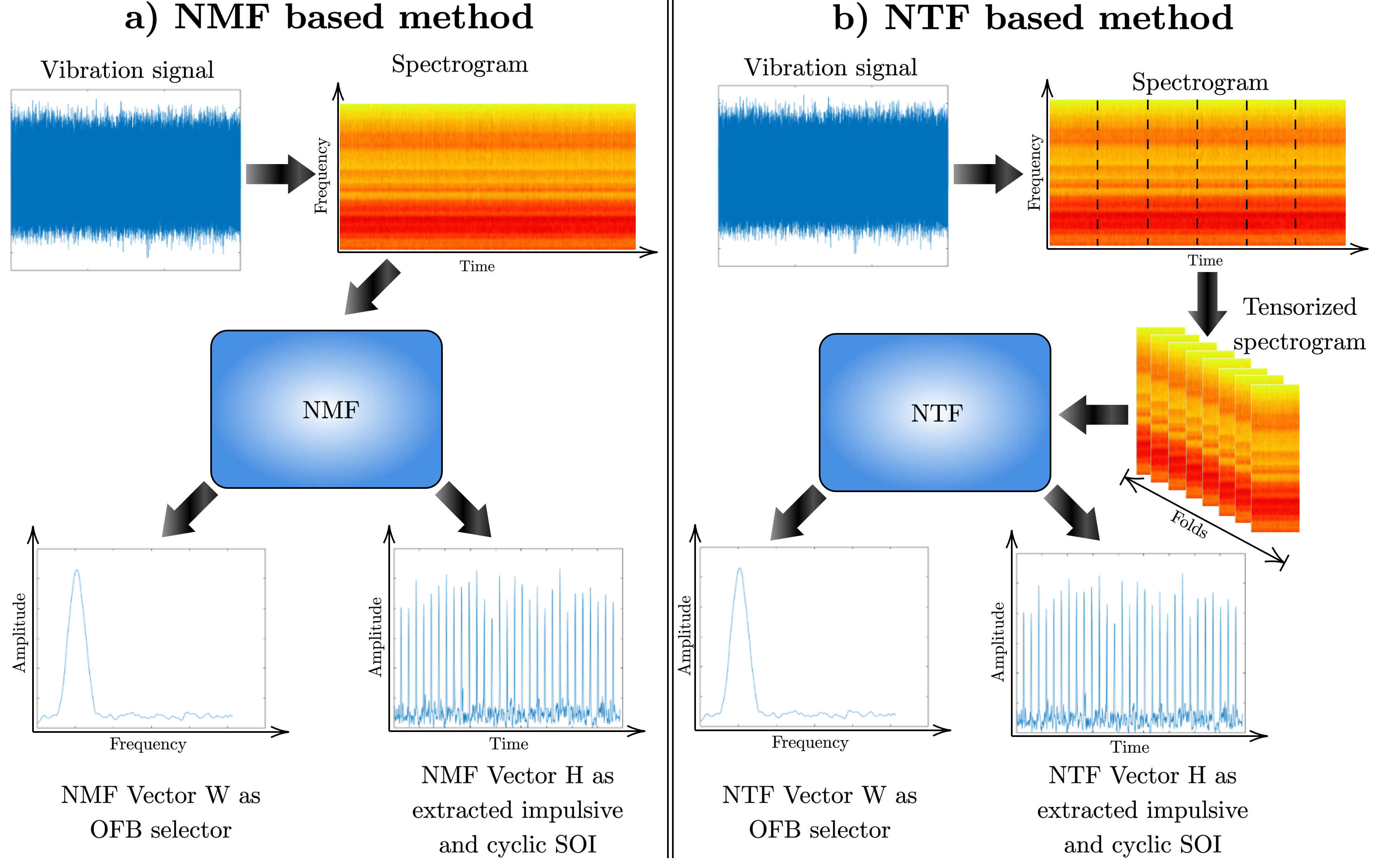} \\
    \caption{Flowcharts of proposed methods: a) NMF-based method, b) NTF-based method.}
    \label{fig:flowcharts}
\end{figure}

For the NMF method, the 1-second excerpt of the mixed signal is taken, and then the spectrogram is calculated. A signal of such a length contains several dozen cyclic impulses, which is enough to identify the period of the cycle that corresponds to the so-called fault frequency. On the spectrogram matrix, NMF is applied, resulting in two matrices: $\bW$ corresponding to the frequency features and $\bH$ corresponding to the time features, which are also referred to as the frequency and time profiles, respectively. This process is illustrated in Figure \ref{fig:flowcharts} (a).

In the case of the NTF method, the entire mixed signal is taken, and then divided into $n$ folds. The spectrogram is calculated for each fold, and then the folds (spectrogram matrices that are 2-way arrays) are arranged into a 3-way array. In the end, the resulting array is factorized using NTF into three loading matrices $\bW$, $\bH$, $\bV$, where $\bW$ and $\bH$ correspond to frequency and time characteristics, respectively, and matrix $\bV$ contains the weights of the folds, but in this case, it is not used for further analysis. The NTF method is shown in Figure \ref{fig:flowcharts} (b).

Due to the intrinsic factor permutation ambiguity of NMF and NTF, the order of the components is randomly assigned each time. To select the most informative one, we applied the skewness metrics to all time profiles, and we took the component with the highest skewness as the best, representing our SOI.

To evaluate the impulsive and cyclic behavior of SOI, we used a modified version of the envelope spectrum based indicator (ENVSI) \cite{hebda2020selection} on time profiles, which could be considered here as spectrum based indicator (SBI).
The SBI is defined as: 
\begin{equation}
\label{SBI_eq}
    SBI = \frac{\sum^{M_1}_{i=1}AIS_i^2}{\sum^{M_2}_{k=1}S_k},
\end{equation}
where $AIS_i$ is the $i$-th magnitude of the SOI in the frequency domain, $S_k$ is the magnitude in the $k$-th frequency bin of the spectrum of the time profile, $M_1$ is the number of harmonics to be analyzed (assuming the SOI is a periodic signal), and $M_2$ determines the number of frequency bins to calculate the total energy. When there are no impulsive components in the time profile, then SBI converges to zero. SBI has a larger value when the impulses in the time profile are stronger (which corresponds to the amplitudes in the spectrum) and the noise is weaker. In our experiments, the number of harmonics $M_1$ is set to 6.

\section{Experiments}
\label{sec:experiments}
There are two types of experiments in this manuscript. The first ones were performed on simulated signals with different SNRs. The purpose of this study was to analyze the efficiency of NMF and NTF to separate the noisy mixed signals generated according to the real-based generative model. SNR was controlled by the variance of the noise. The aim is to determine the range of SNR for which both methods can extract the SOI with the assumed quality. For the best SNR, the SOI was clearly seen in the observation. For the worst SNR, the cyclic impulsive signal was not visible at all, even on the spectrogram.

The second type of experiments were performed on a real vibration signal measured on the test rig. One of the bearings was intentionally damaged; however, cyclic impulses were not visible in the time domain. An expert with the appropriate tools and experience is able to detect the presence of the SOI (damage) in the spectrogram.

The ultimate goal is to apply NTF to a vibration signal with local damage and present the ability of NTF to detect faults. However, to highlight the superiority of NTF, we refer to the NMF approach which has already been analyzed as a powerful tool for local damage detection.

As introduced in section \ref{sec:algorithm}, the multiplicative update rules were used to estimate the factors of NMF and NTF with three different cost functions that come from the $\beta$-divergence. To make a fair comparison, the same settings were used for NMF and NTF:  decomposition rank $J$ was set to two, the maximum number of iterations to 100, and the drop in the normalized residual error for subsequent iterations was below $10^{-5}$. All the spectrograms were computed using the Hamming window of the length equal to 128 was, the overlap set to 100 samples, and the number of DFT points set to 512.

\subsection{Simulated signals}
In order to assess decomposition capability of NMF and NTF, we generated 26 mixed signals of 60 seconds duration with different SNR values, ranging from -5.84 to -21.39 dB, with a sampling rate of 25 kHz. The SNR is defined as:
\begin{equation}
    SNR = 20 \log_{10} \frac{||s(t)||_2}{||(d(t)+n(t))||_2},
\end{equation}
where $||s(t)||_2$ is the $l_2$ norm of the SOI, $||(d(t)+n(t))||_2$ is the $l_2$ norm of the noisy disturbances (non-cyclic impulsive perturbations $d(t)$ and Gaussian noise $n(t)$). Negative values of SNRs mean that the signal power is below the noise power. The simulated signals represent the vibration signals of bearing damage from the crushing machine. The SOI was generated according to model (\ref{eq_2}) with the amplitude set to 3 and  the local fault frequency equal to 30 Hz ($1/T_p$). The non-cycle impulses were also modulated but their localization was random, and their amplitude was set to 5. The total number of non-cycle impulses amounts to 30. The carrier frequency ($f_c$) of cyclic and non-cyclic impulses is 2500 Hz and 6000 Hz, respectively. The standard deviation of white noise in the simulated signals changed in the range from 0.5 to 3.

\begin{figure}[H]
    \centering
    \includegraphics[scale=0.4]{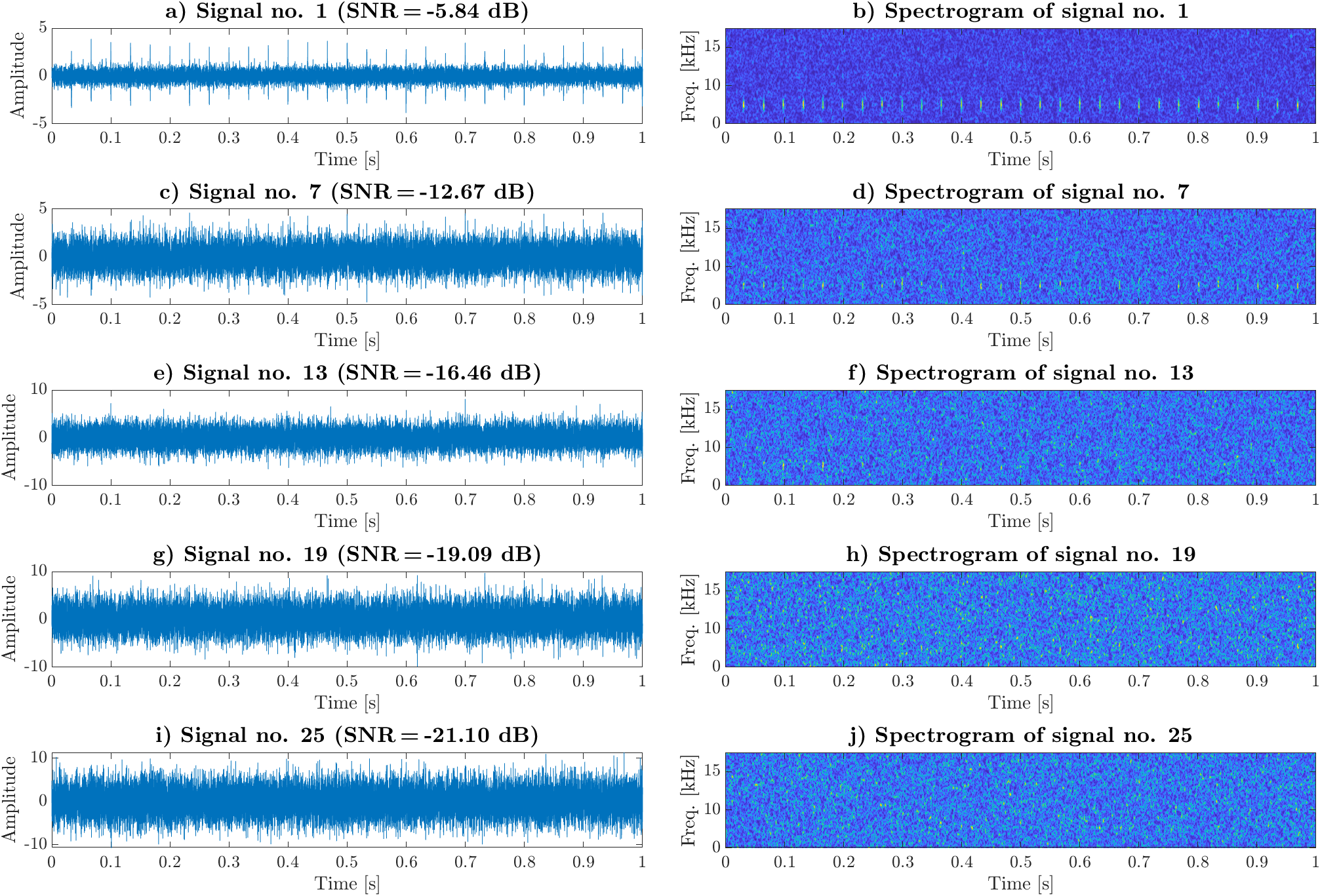}
    \caption{Selected simulated signals and their spectrograms: a) signal no. 1, b) spectrogram of signal no. 1, c) signal no. 7, d) spectrogram of signal no. 7, e) signal no. 13, f) spectrogram of signal no. 13, g) signal no. 19, h) spectrogram of signal no. 19, i) signal no. 25, j) spectrogram of signal no. 25.}
    \label{fig:simulated_signals}
\end{figure}

We took the first second for the NMF-based analysis, and the original mixed signal for the NTF-based analysis was tensorized into 40 folds -- each one second long. The selected simulated signals and their spectrograms are presented in Figure \ref{fig:simulated_signals}.

\subsection{Test rig data}

\begin{figure}[h!]
    \centering
    \includegraphics[width=0.5\textwidth]{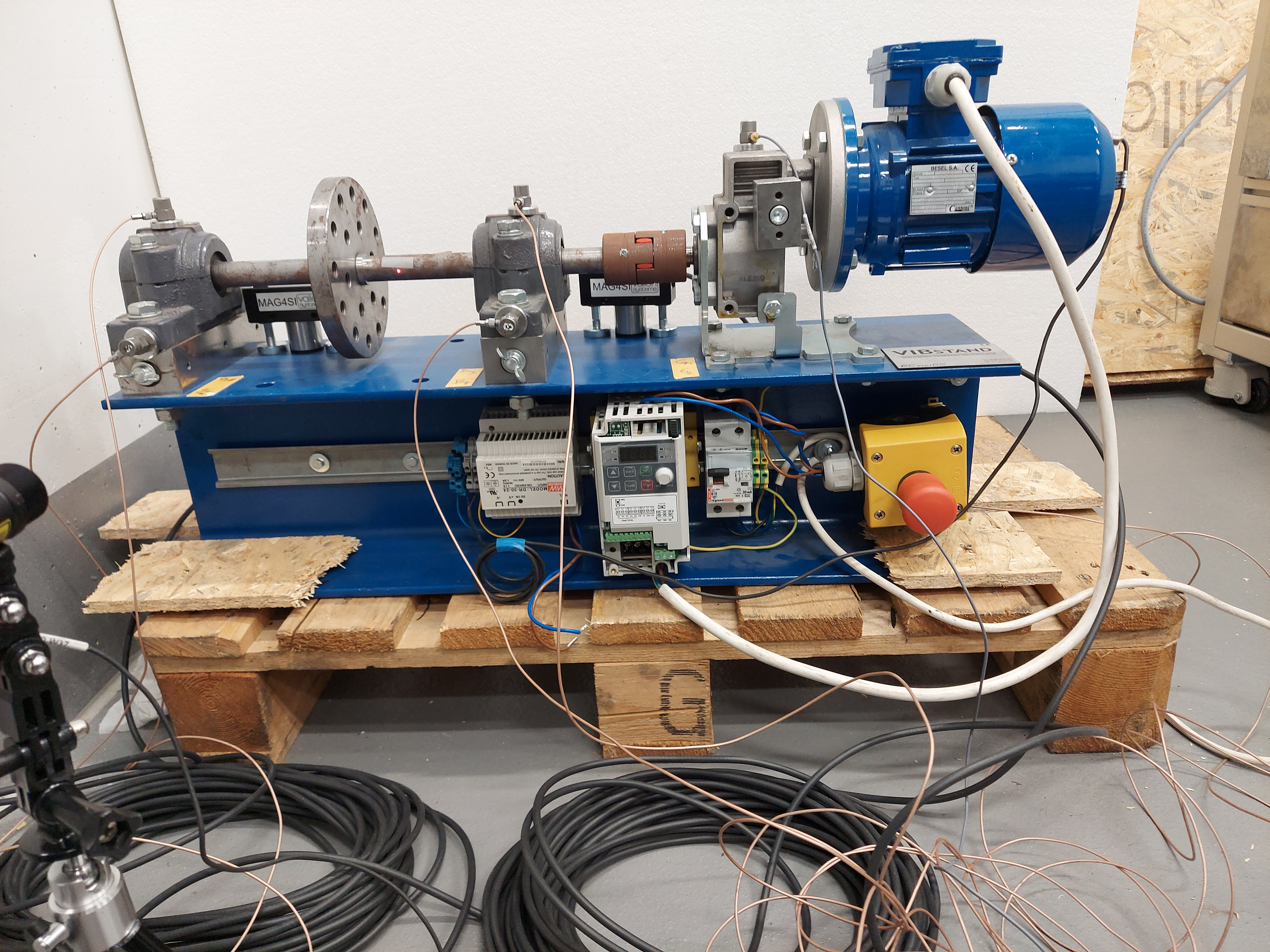}
    \caption{Test rig used in the experiment.}
    \label{fig:rig}
\end{figure}

We used a test rig (presented in Figure \ref{fig:rig}) that contained an electric motor, gearbox, couplings, and two bearings, of which one was deliberately damaged. We recorded a 40-second-long vibration signal with a sampling rate of 50 kHz using an accelerometer (KISTLER Model 8702B500), which was stacked horizontally to the shaft bearing. Similarly to simulations, the first second of signal was used for NMF, and the signal for NTF was divided into 40 folds -- each one-second-long, creating the tensor. 

\begin{figure}[H]
    \centering
    \includegraphics[scale=0.45]{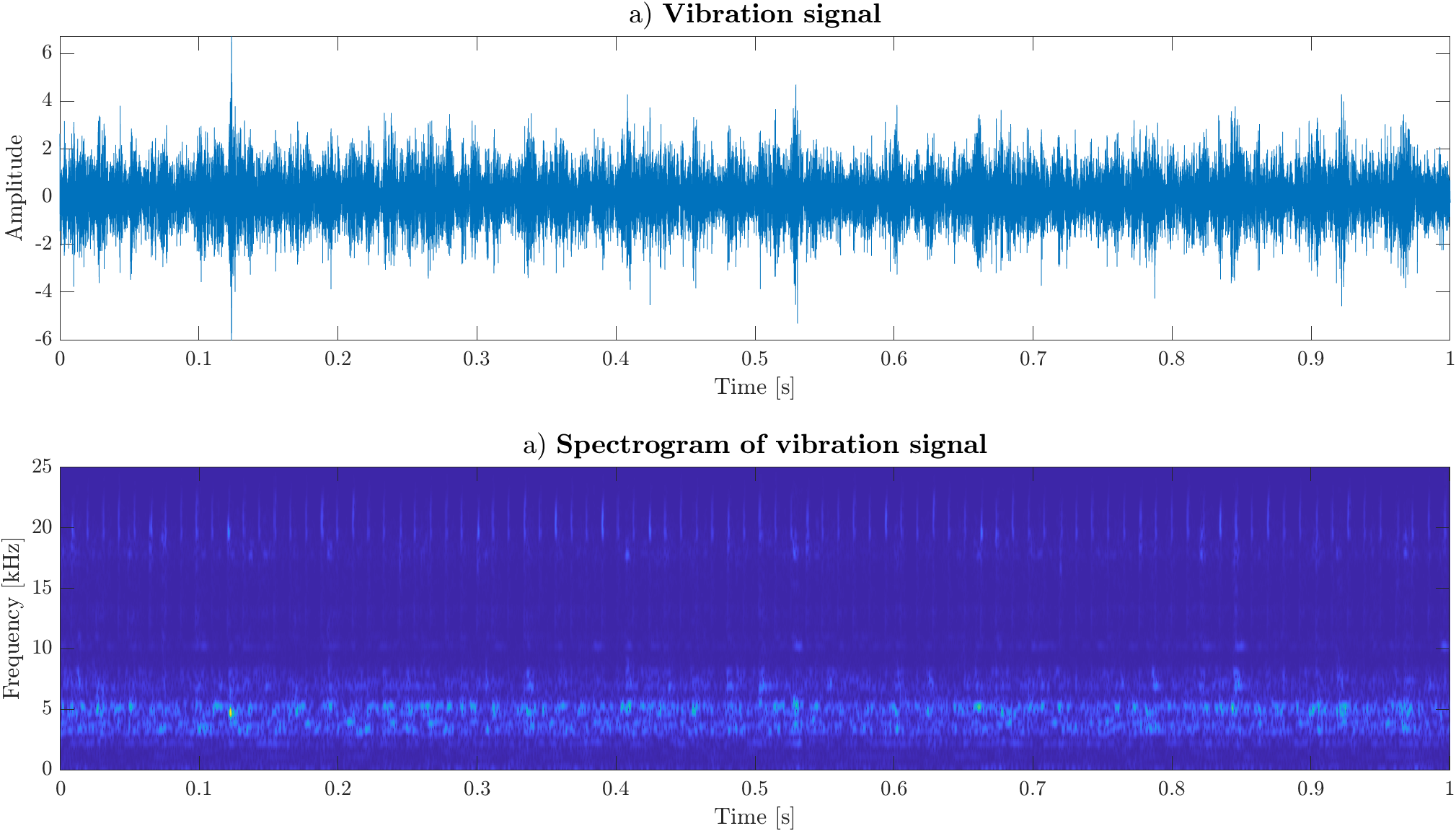}
    \caption{Recorded vibration signal and its spectrogram: a) vibration signal, b) spectrogram of vibration signal.}
    \label{fig:real_signal}
\end{figure}

The first second of the recorded real vibration signal (top panel) and its spectrogram (bottom panel) are shown in Figure \ref{fig:real_signal}.

\section{Results}
\label{sec:results}
\subsection{Results for simulated signals}

\begin{figure}[H]
\centering
    \begin{subfigure}[b]{.325\textwidth}
        \centering
        \includegraphics[width=\textwidth]{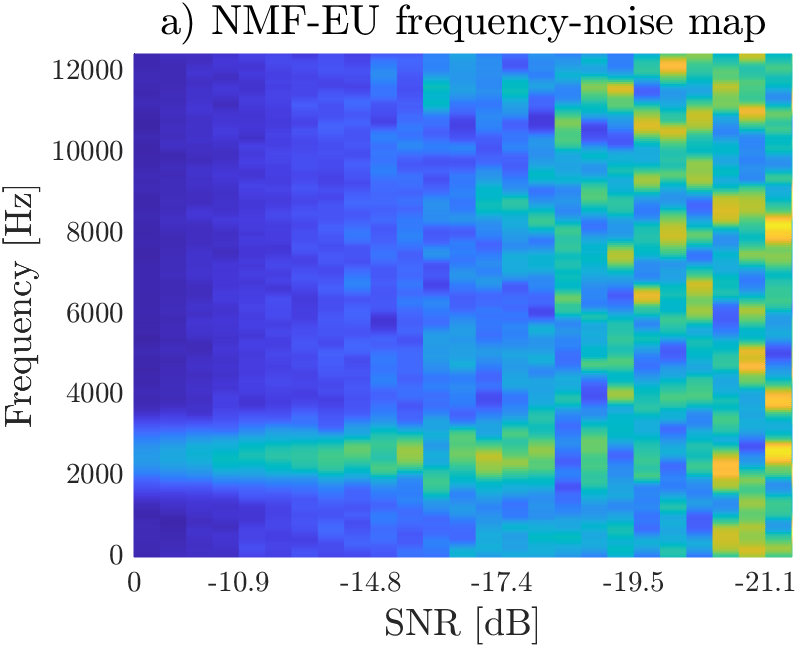}
    \end{subfigure}
    \hfill
    \begin{subfigure}[b]{.325\textwidth}
        \centering
        \includegraphics[width=\textwidth]{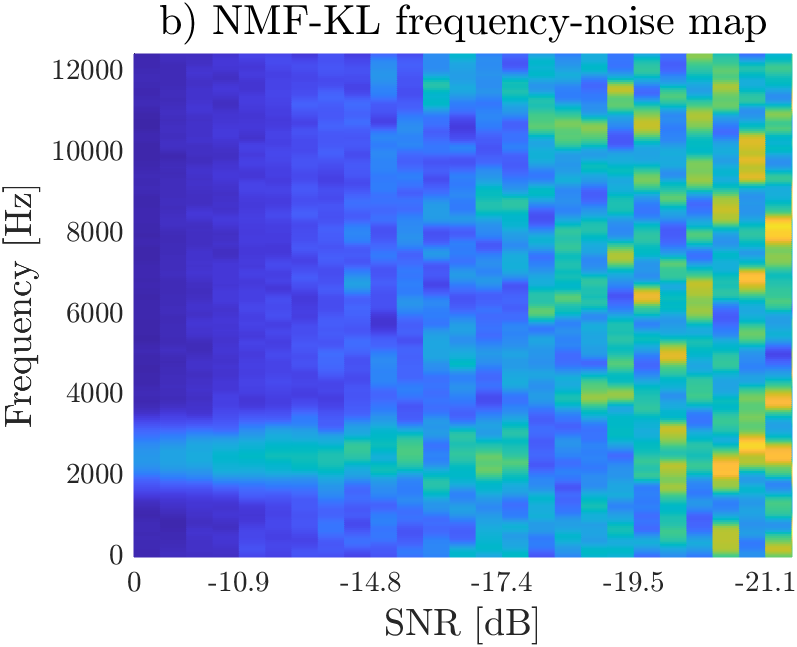}
    \end{subfigure}
    \hfill
    \begin{subfigure}[b]{.325\textwidth}
        \centering
        \includegraphics[width=\textwidth]{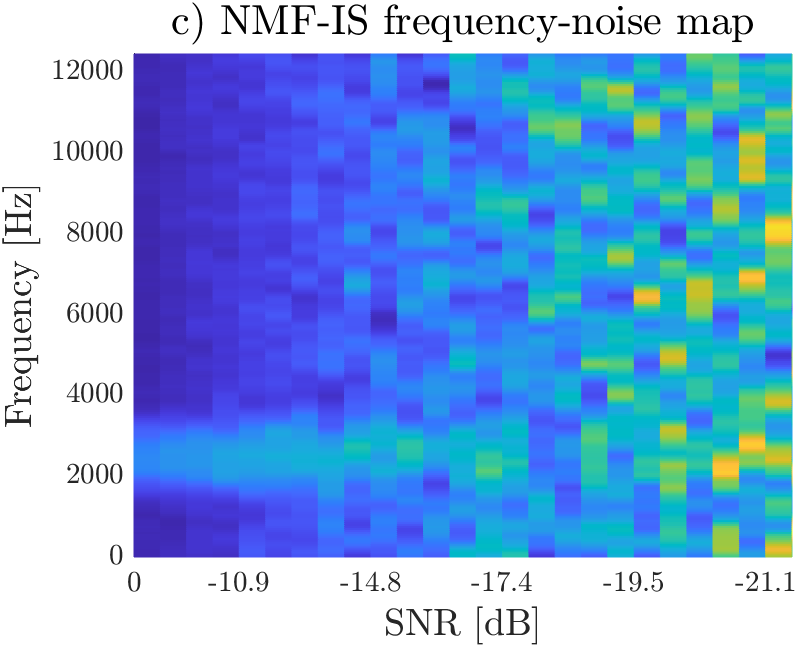}
    \end{subfigure}
    \newline
    \vspace{-5pt}
    \begin{subfigure}[b]{.325\textwidth}
        \centering
        \includegraphics[width=\textwidth]{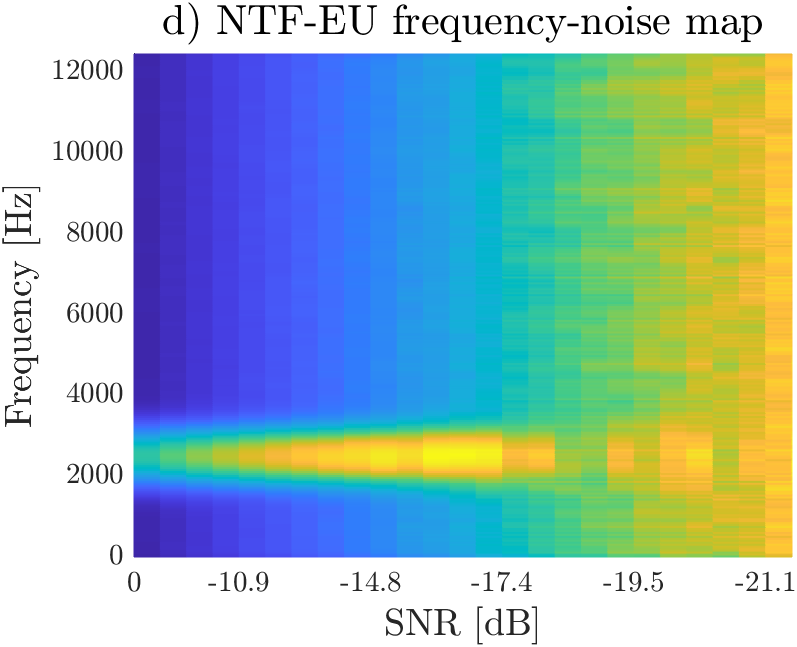}
    \end{subfigure}
    \hfill
    \begin{subfigure}[b]{.325\textwidth}
        \centering
        \includegraphics[width=\textwidth]{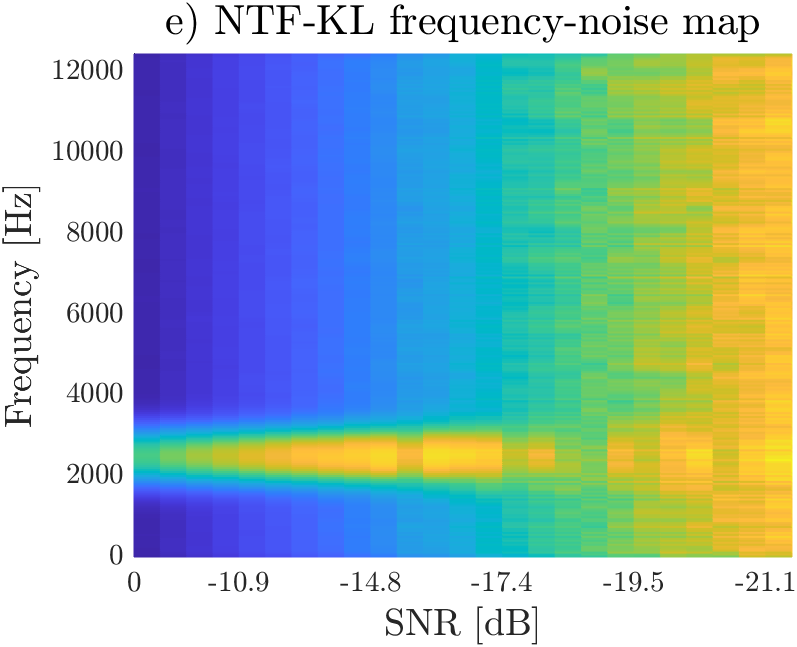}
    \end{subfigure}
    \hfill
    \begin{subfigure}[b]{.325\textwidth}
        \centering
        \includegraphics[width=\textwidth]{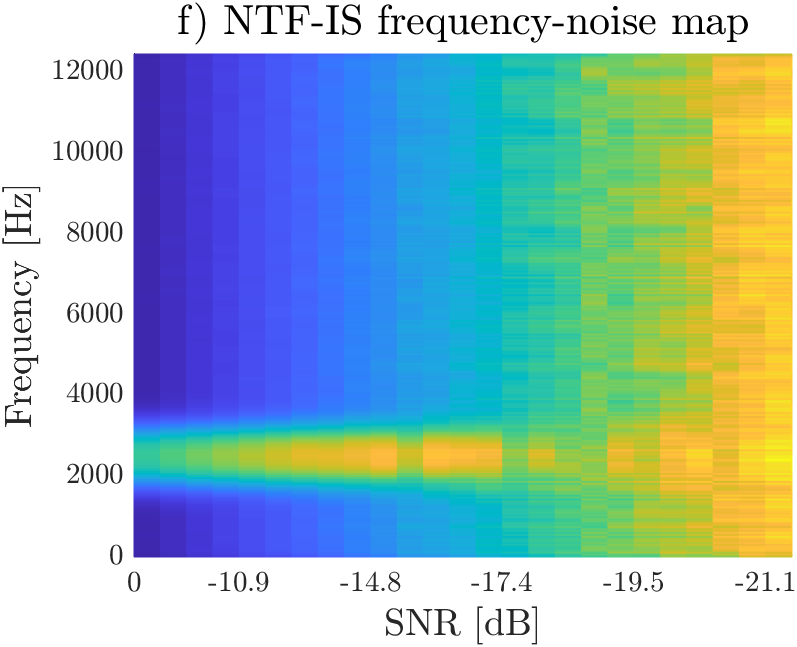}
    \end{subfigure}
    
    \caption{NMF and NTF frequency-noise maps from simulated signals: a) NMF-EU, b) NMF-KL, c) NMF-IS, d) NTF-EU, e) NTF-KL, f) NTF-IS.}
    \label{best_freq}
\end{figure}

\begin{figure}[H]
\centering
    \begin{subfigure}[b]{.325\textwidth}
        \centering
        \includegraphics[width=\textwidth]{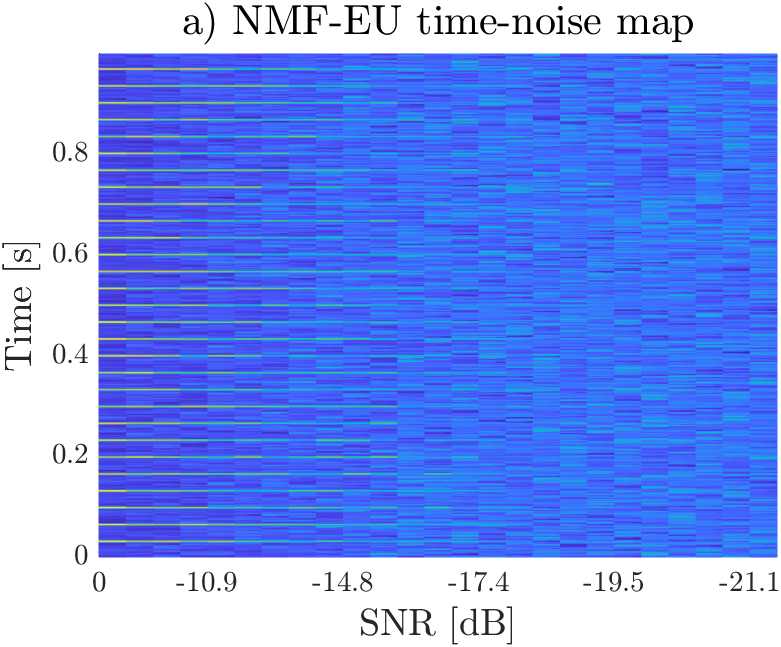}
    \end{subfigure}
    \hfill
    \begin{subfigure}[b]{.325\textwidth}
        \centering
        \includegraphics[width=\textwidth]{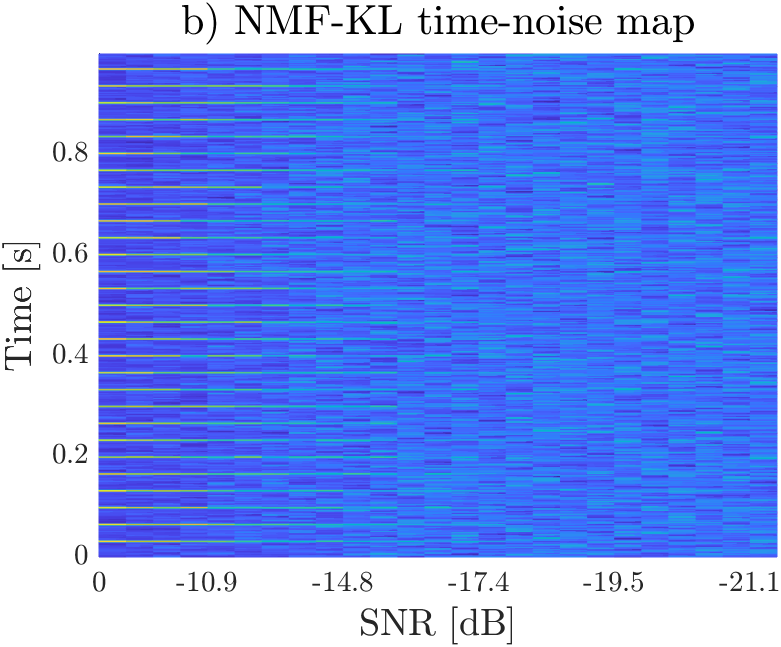}
    \end{subfigure}
    \hfill
    \begin{subfigure}[b]{.325\textwidth}
        \centering
        \includegraphics[width=\textwidth]{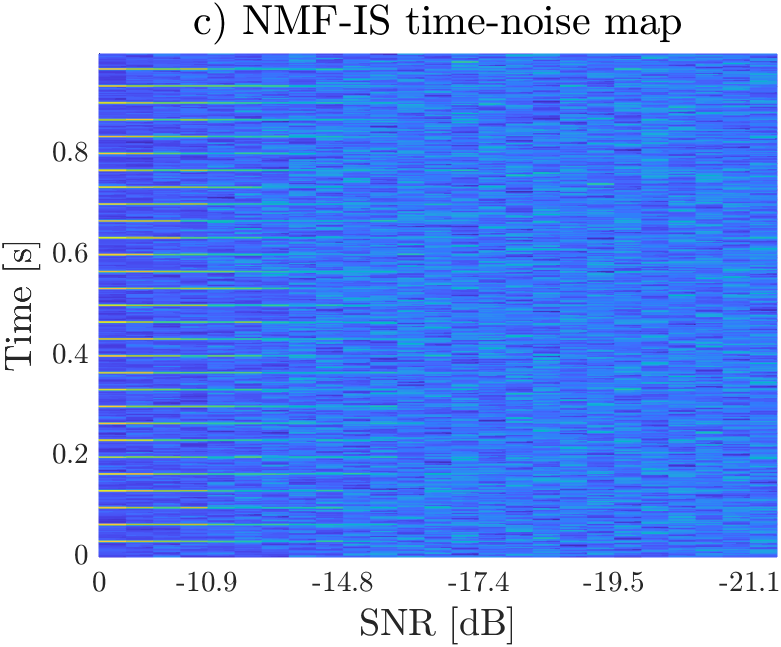}
    \end{subfigure}
    \newline
    \vspace{-5pt}
    \begin{subfigure}[b]{.325\textwidth}
        \centering
        \includegraphics[width=\textwidth]{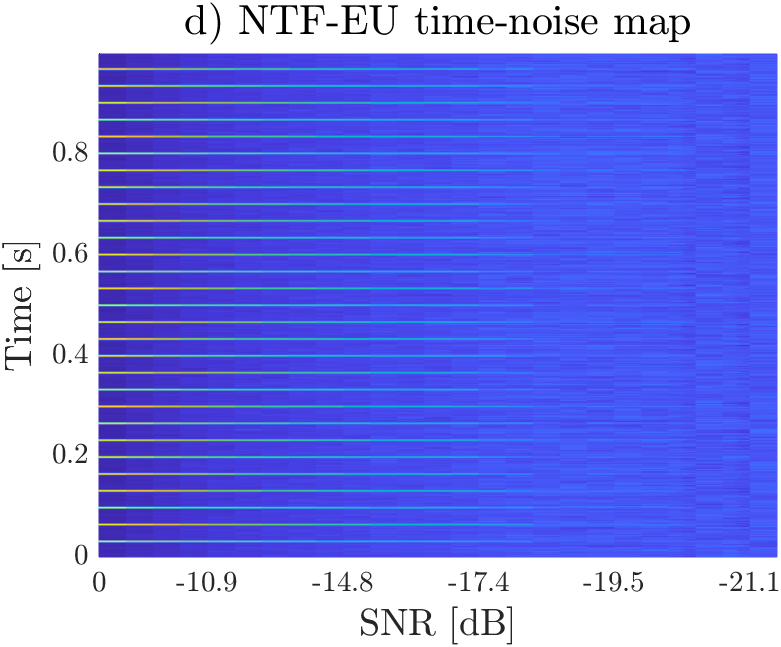}
    \end{subfigure}
    \hfill
    \begin{subfigure}[b]{.325\textwidth}
        \centering
        \includegraphics[width=\textwidth]{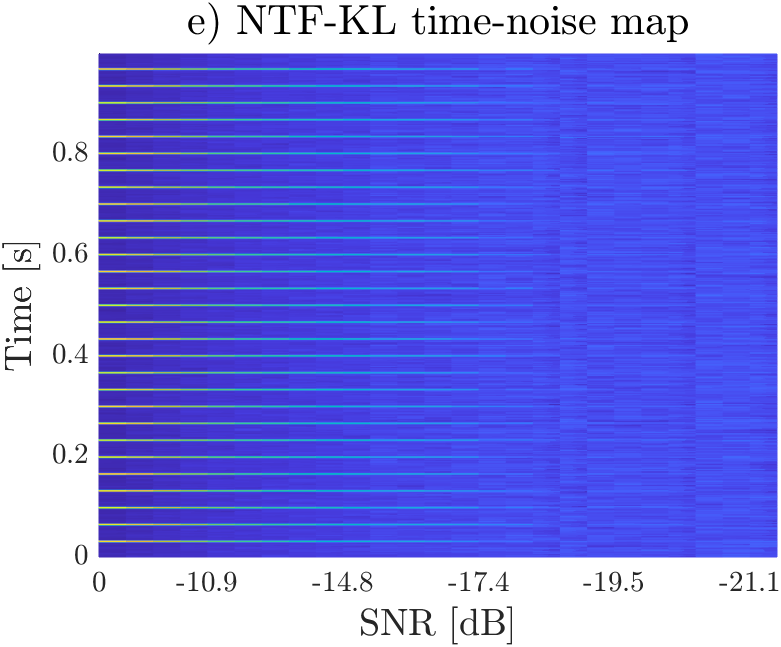}
    \end{subfigure}
    \hfill
    \begin{subfigure}[b]{.325\textwidth}
        \centering
        \includegraphics[width=\textwidth]{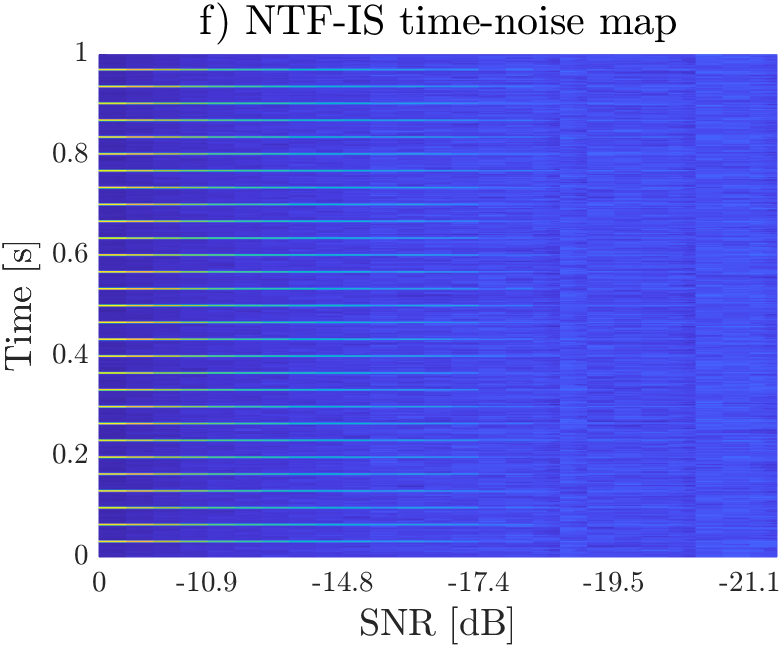}
    \end{subfigure}
    
    \caption{NMF and NTF time-noise maps from simulated signals: a) NMF-EU, b) NMF-KL, c) NMF-IS, d) NTF-EU, e) NTF-KL, f) NTF-IS.}
    \label{best_time}
\end{figure}

The results obtained by the factorization methods (in the diagnostic context) are presented in Figures \ref{best_freq} and \ref{best_time}. Figure \ref{best_freq} illustrates the ability to identify the optimal frequency band (OFB) for each simulation case -- the simulation signal taken from a group of 26 mixed signals. The OFB is located around 2500 Hz. Factorizing the spectrogram, we receive frequency feature vector $\bw_c$ that highlights the presence of the SOI (see Figure \ref{fig:flowcharts}). As it is repeated for each simulation case with decreasing SNR, we obtain a matrix that can be regarded as a frequency-noise map. In Figure \ref{best_freq}, we present the frequency noise maps for three cost functions (EU, KL, and IS) for both the NMF and NTF methods. It could be seen that NMF is losing the ability to point out the OFB at the SNR around -15dB. For smaller SNRs, the importance of noise components increases. For the NTF case, such a problem appears much later, i.e., up to -20 dB we are still able to notice the OFB. 

Similarly, Figure \ref{best_time} illustrates the time-noise maps that demonstrate the change of the corresponding time feature vector $\bh_c$ versus the SNR. Again, we present 6 time-noise maps: for 3 cost functions and for both NMF and NTF methods. The maps allow us to determine the range of SNR for which an impulsive behavior of the SOI can be observed.

\begin{figure}[H]
    \centering
    \includegraphics[scale=0.5]{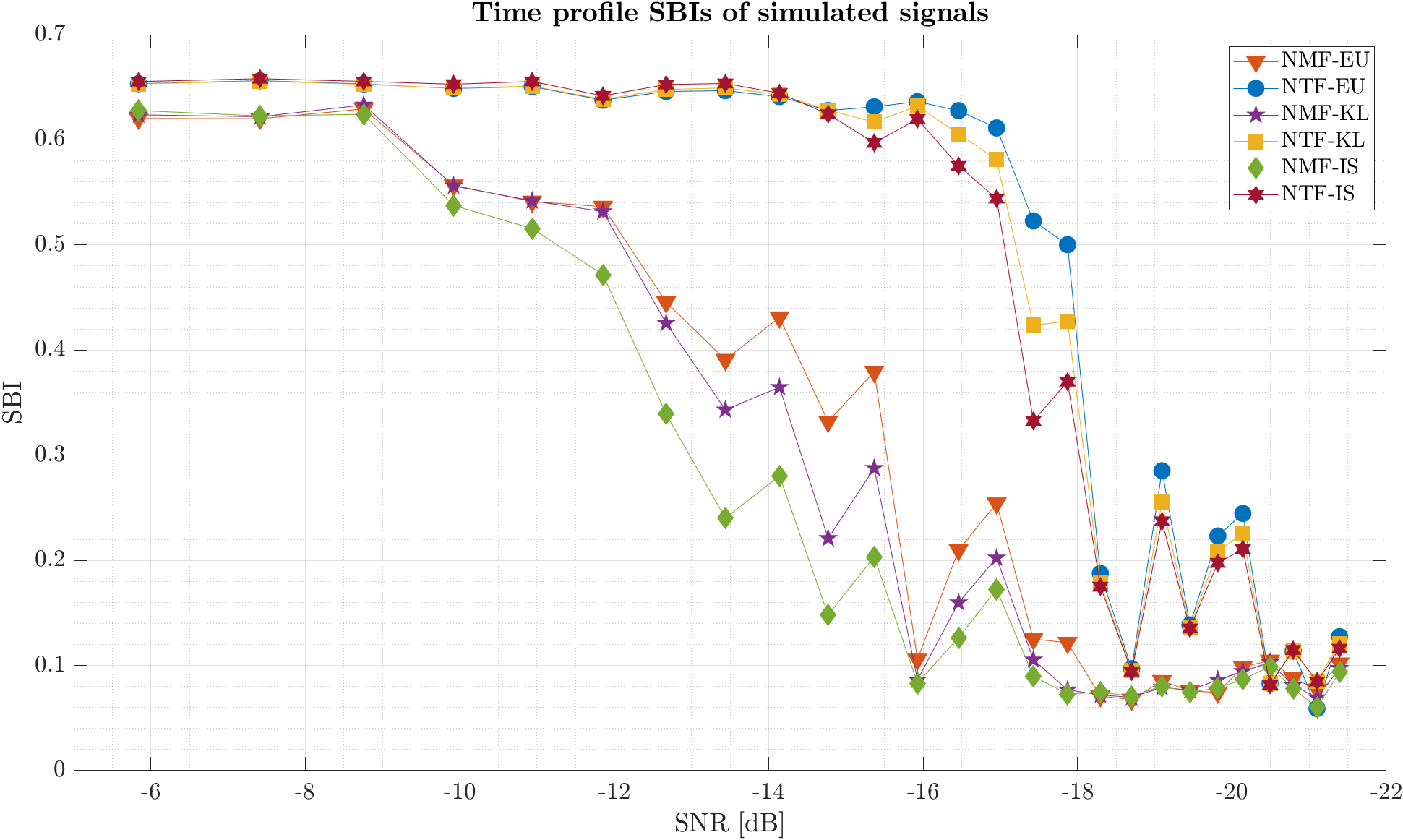}
    \caption{SBIs of {extracted time profiles} for each method.}
    \label{fig:sbi}
\end{figure}

Figure \ref{fig:sbi} summarizes our finding using the proposed SBI measure. The plots show a decrease in SBI versus a decrease in SNR for all the methods evaluated.  As can be seen, it is clear that for all the cost functions, NTF considerably outperforms NMF. The decrease in SBI is much slower for NTF than for NMF. From the frequency-noise maps, it can be noted that IFB for NTF is clearly detectable, and this is not the case for NMF. NMF is unable to detect IFB and provides a strong background noise. Furthermore, the results presented in the time-noise maps show that the impulsive component of the signals is visible for lower SNRs, and similarly to the frequency-noise maps, the level of background noise is lower for NTF than for NMF.

When we compare the cost functions, the simulations show that there is no large difference between the results obtained by the discussed cost functions, but the EU distance seems to achieve the result with the lowest drop in SBI.

\subsection{Results for real vibration data}
\begin{figure}[H]
\centering

    \begin{subfigure}[b]{.325\textwidth}
        \centering
        \includegraphics[width=\textwidth]{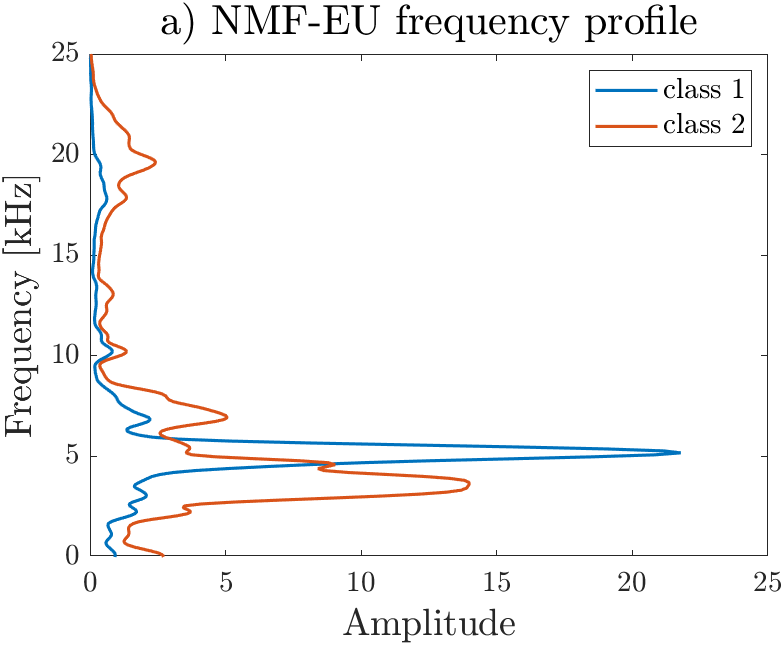}
    \end{subfigure}
    \hfill
    \begin{subfigure}[b]{.325\textwidth}
        \centering
        \includegraphics[width=\textwidth]{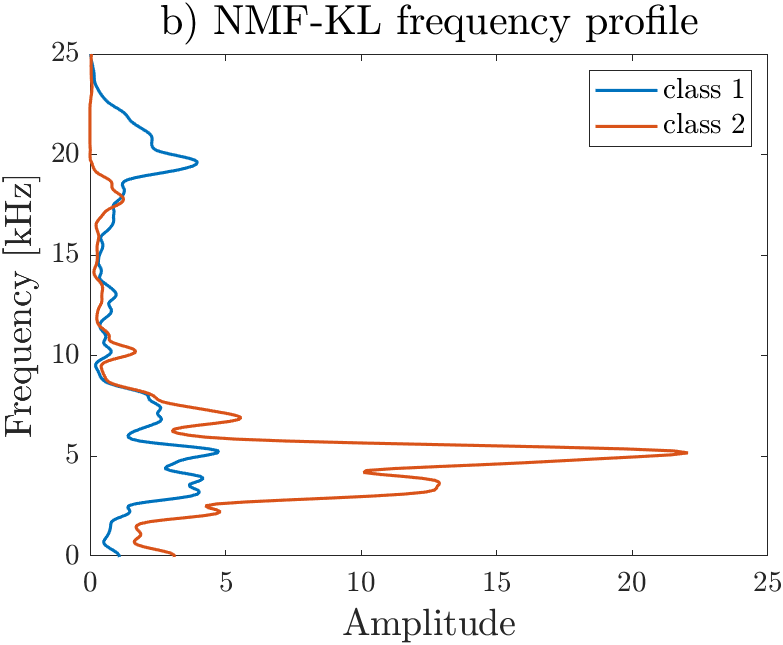}
    \end{subfigure}
    \hfill
    \begin{subfigure}[b]{.325\textwidth}
        \centering
        \includegraphics[width=\textwidth]{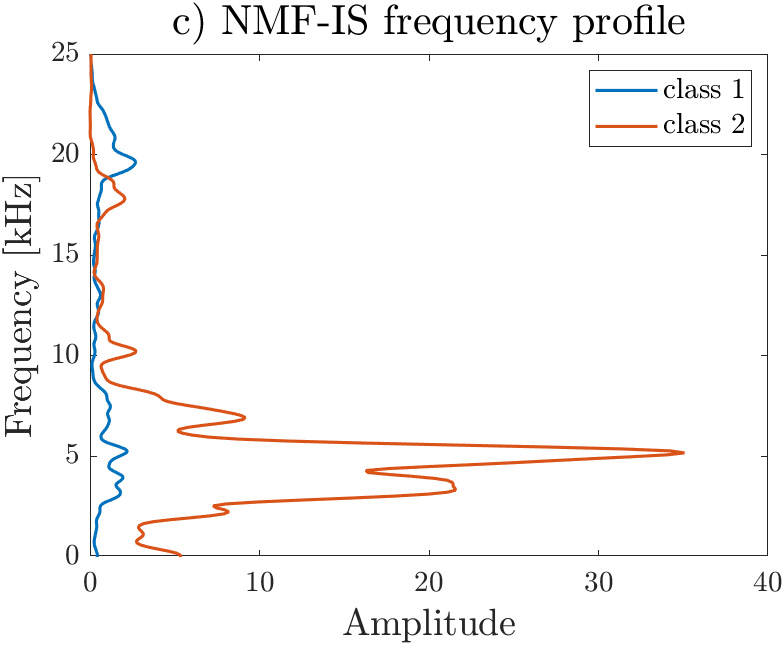}
    \end{subfigure}
    \newline
    
    \begin{subfigure}[b]{.325\textwidth}
        \centering
        \includegraphics[width=\textwidth]{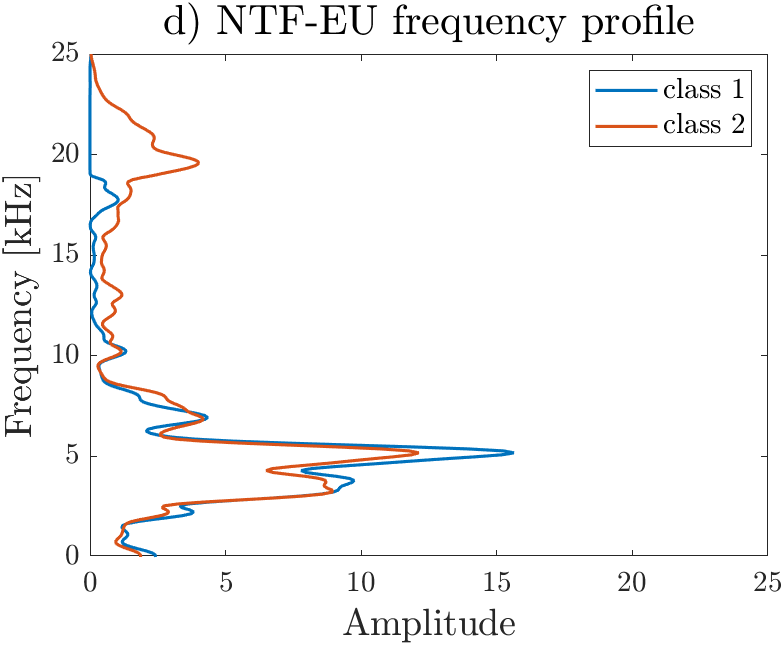}
    \end{subfigure}
    \hfill
    \begin{subfigure}[b]{.325\textwidth}
        \centering
        \includegraphics[width=\textwidth]{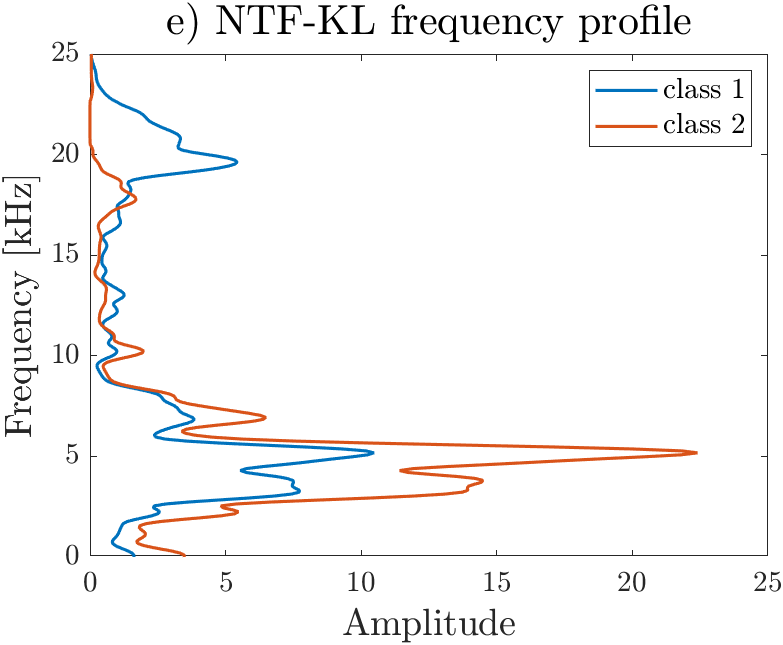}
    \end{subfigure}
    \hfill
    \begin{subfigure}[b]{.325\textwidth}
        \centering
        \includegraphics[width=\textwidth]{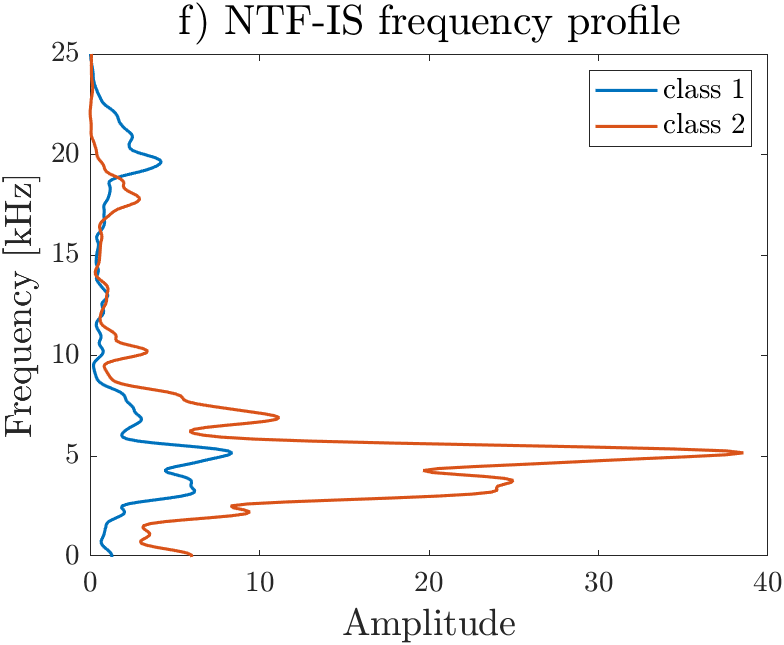}
    \end{subfigure}
    
    \caption{NMF and NTF frequency profiles of real vibration signal: a) NMF-EU, b) NMF-KL, c) NMF-IS, d) NTF-EU, e) NTF-KL, f) NTF-IS.}
    \label{fig:nmf_ntf_real_freq_euclidean}
\end{figure}

\begin{figure}[H]
\centering

    \begin{subfigure}[b]{.325\textwidth}
        \centering
        \includegraphics[width=\textwidth]{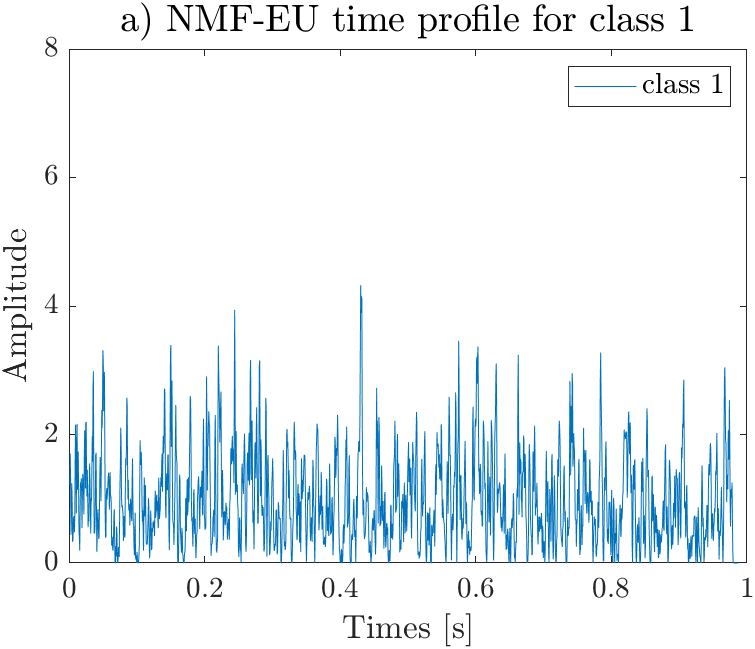}
    \end{subfigure}
    \hfill
    \begin{subfigure}[b]{.325\textwidth}
        \centering
        \includegraphics[width=\textwidth]{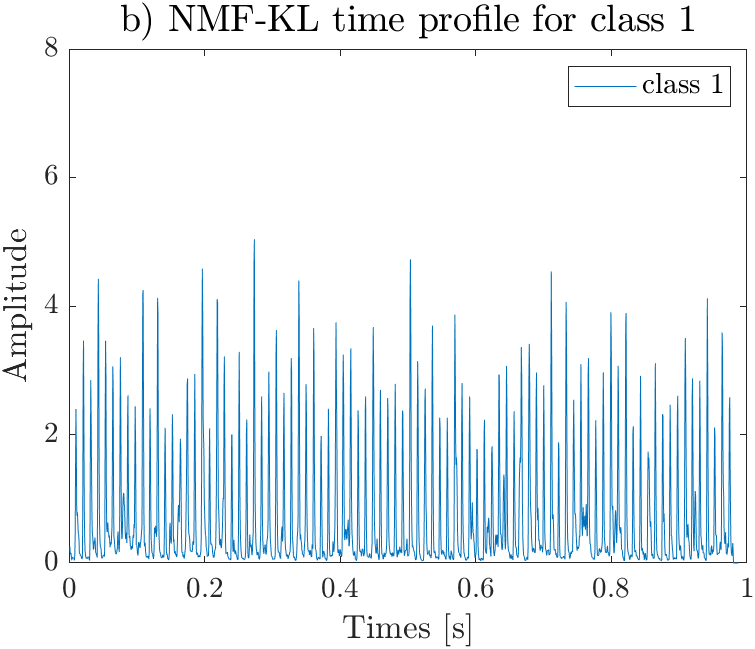}
    \end{subfigure}
    \hfill
    \begin{subfigure}[b]{.325\textwidth}
        \centering
        \includegraphics[width=\textwidth]{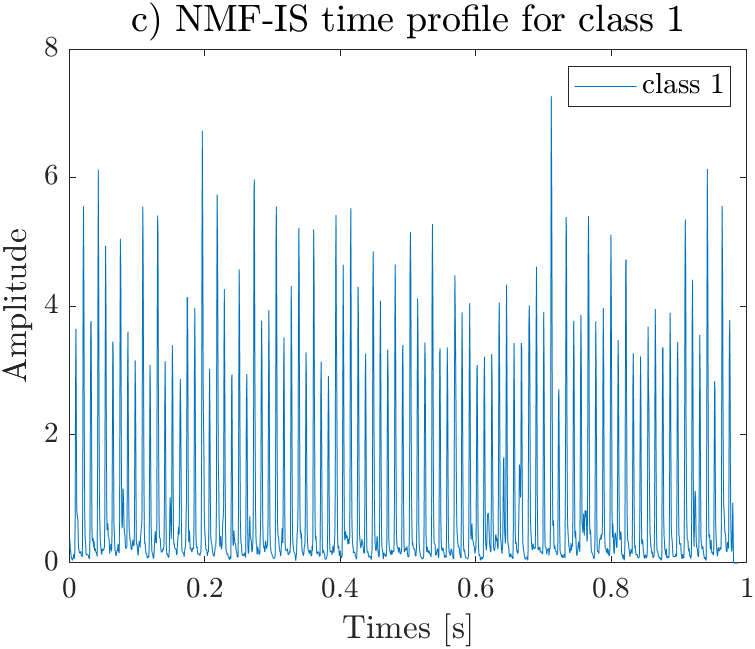}
    \end{subfigure}
    \newline
    \vspace{-5pt}
    \begin{subfigure}[b]{.325\textwidth}
        \centering
        \includegraphics[width=\textwidth]{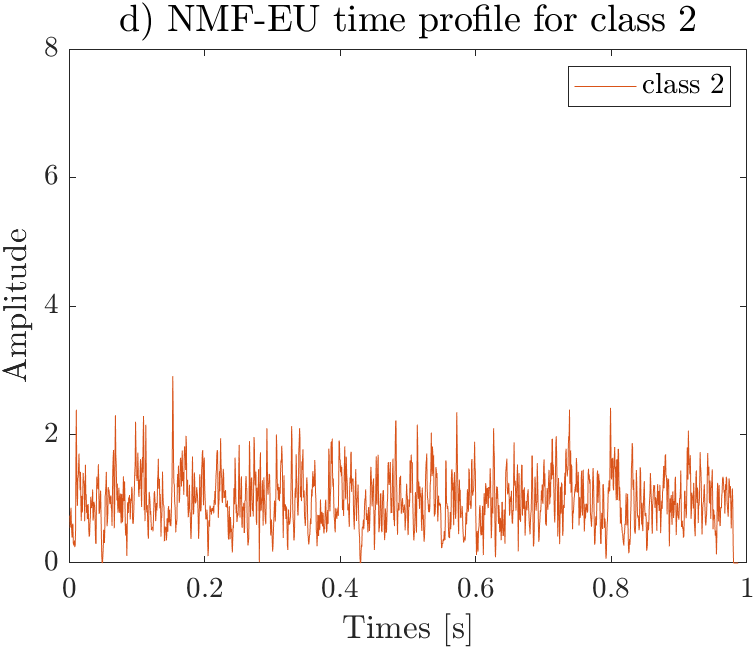}
    \end{subfigure}
    \hfill
    \begin{subfigure}[b]{.325\textwidth}
        \centering
        \includegraphics[width=\textwidth]{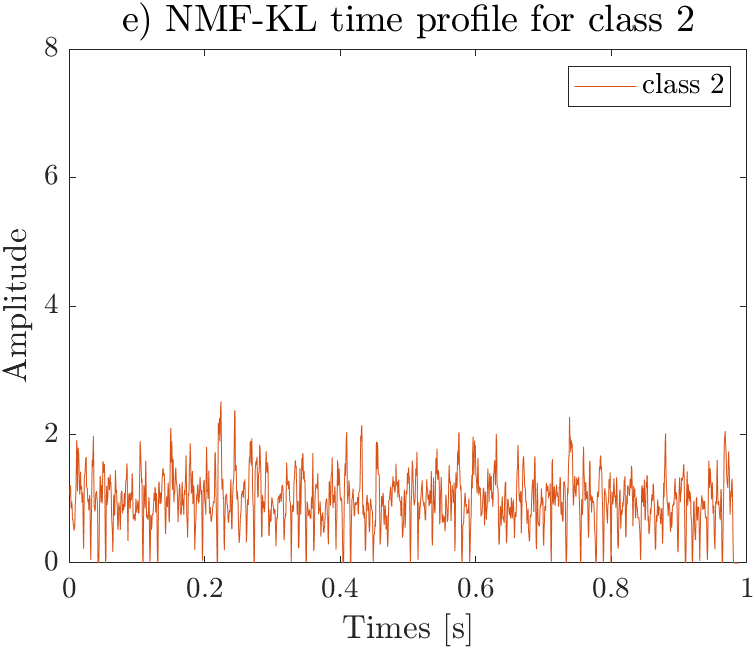}
    \end{subfigure}
    \hfill
    \begin{subfigure}[b]{.325\textwidth}
        \centering
        \includegraphics[width=\textwidth]{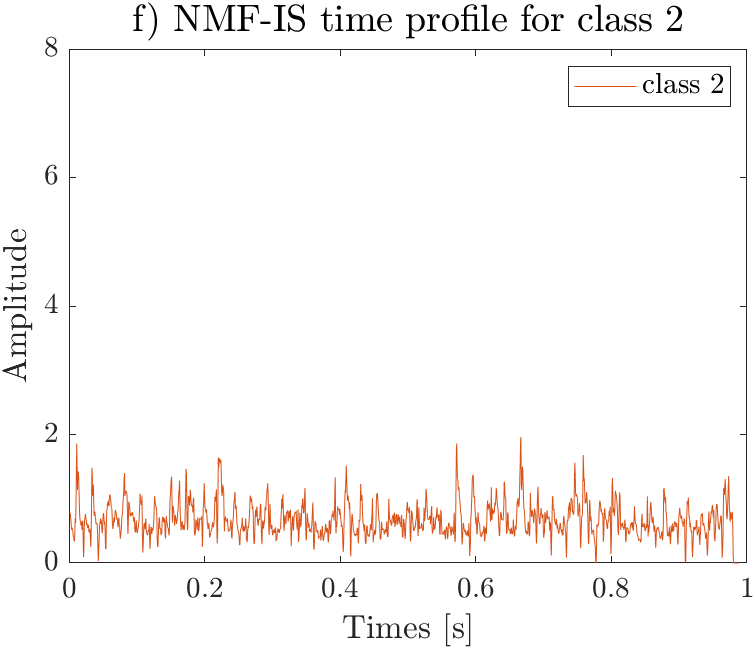}
    \end{subfigure}
    \newline
    \vspace{-5pt}
    \caption{NMF time profiles of real vibration signal: a) NMF-EU class 1, b) NMF-KL class 1, c) NMF-IS class 1, d) NMF-EU class 2, e) NMF-KL class 2, f) NMF-IS class 2.}
    \label{fig:nmf_time_profiles}
\end{figure}

In this section, we analyze the results obtained in the experiments with a real vibration signal. Figure \ref{fig:nmf_ntf_real_freq_euclidean} presents the frequency profiles obtained from the NMF and NTF methods. In turn, Figure \ref{fig:nmf_time_profiles} presents the NMF and NTF time profiles for two classes with different cost functions.

From the frequency profiles, one may conclude that both NMF and NTF point out the resonance around 4-5 kHz. From the spectrogram, we cannot see an impulsive and periodic behavior. Another interesting area in the frequency profile is located around 20 kHz. The results for NMF gently highlight this area (mostly blue curves – class 1 – for NMF-KL and NMF-IS). In the case of NTF, the curve for class 2 (NTF-EU) and the curves for class 1 (both NTF-KL and IS) point out this area.

Since the frequency profiles do not provide the reliable information on the OFB localization, the time profiles should also be analyzed. The impulsive and periodic nature of the time profile in conjunction with the information captured from the frequency profile could be regarded as a reliable source of information on the occurrence of local damage.


Looking at the time profiles, it can be noted that the class one in NMF-EU does not yield the reliable information on an impulse and period behavior. However, a minor success can be observed for the other two cost functions, especially for the IS distance. Class 2 does not provide any useful information for all the cost functions. 
For the NTF method, the first class is generally more informative than the second one.
For the NMF-EU, the difference between class 1 and class 2 is not obvious, the extracted signal seems to have a periodic and impulsive nature, but it is much worse than for other criteria. For both NTF-KL and NTF-IS, class 1 is much more informative than class 2. The extracted signals are clearly impulsive and cyclic. The best results are obtained for NTF-IS as the heights of the impulses are larger than those for the other criteria.

\begin{figure}[H]
\centering

    \begin{subfigure}[b]{.325\textwidth}
        \centering
        \includegraphics[width=\textwidth]{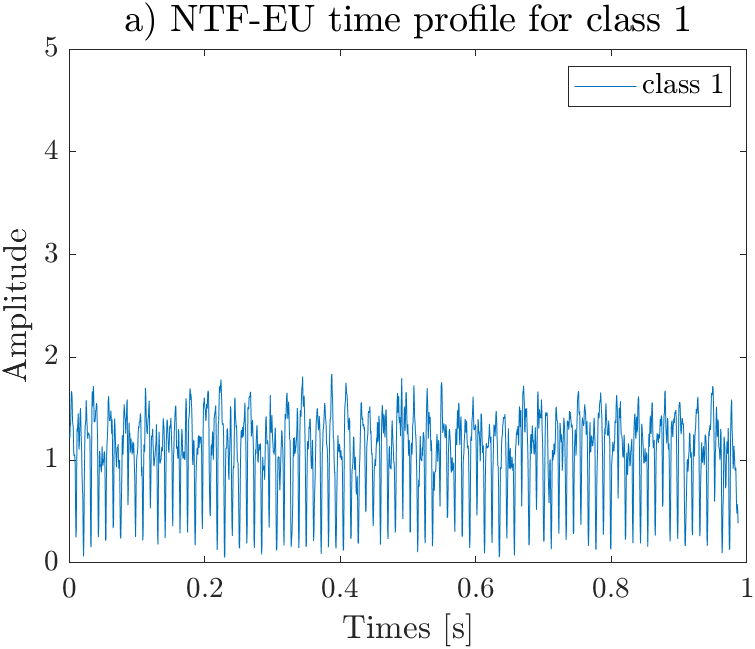}
    \end{subfigure}
    \hfill
    \begin{subfigure}[b]{.325\textwidth}
        \centering
        \includegraphics[width=\textwidth]{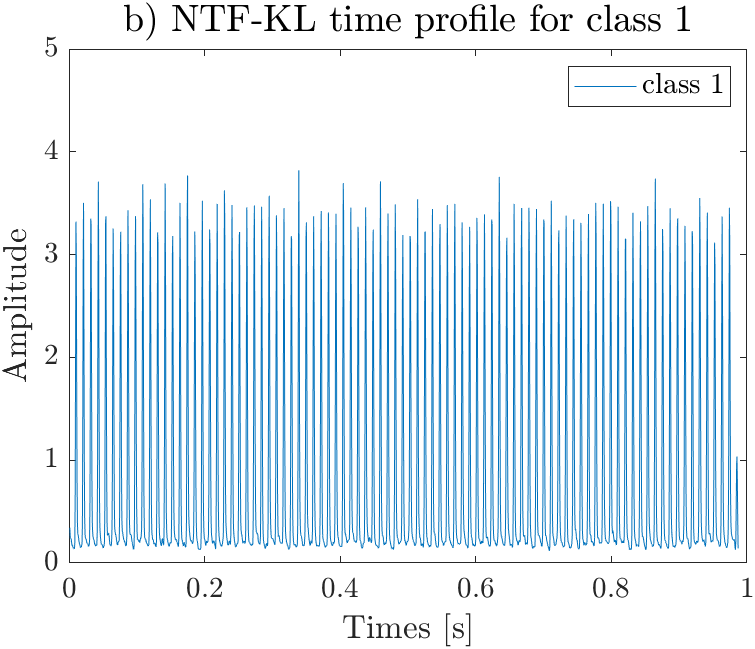}
    \end{subfigure}
    \hfill
    \begin{subfigure}[b]{.325\textwidth}
        \centering
        \includegraphics[width=\textwidth]{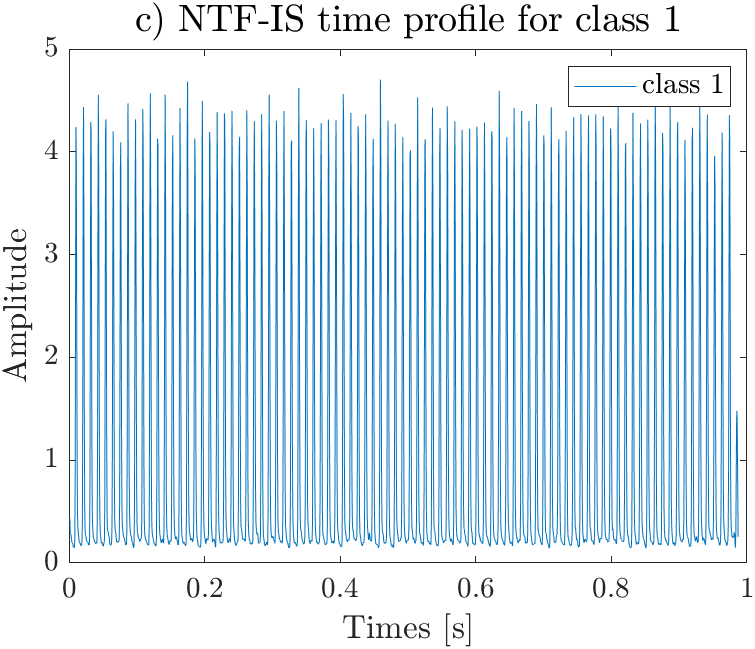}
    \end{subfigure}
    \newline
    \vspace{-5pt}
    \begin{subfigure}[b]{.325\textwidth}
        \centering
        \includegraphics[width=\textwidth]{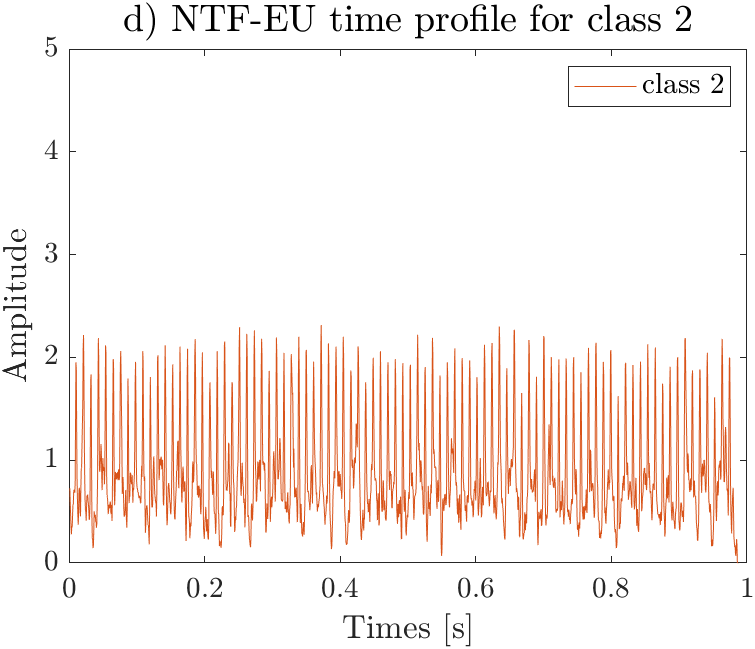}
    \end{subfigure}
    \hfill
    \begin{subfigure}[b]{.325\textwidth}
        \centering
        \includegraphics[width=\textwidth]{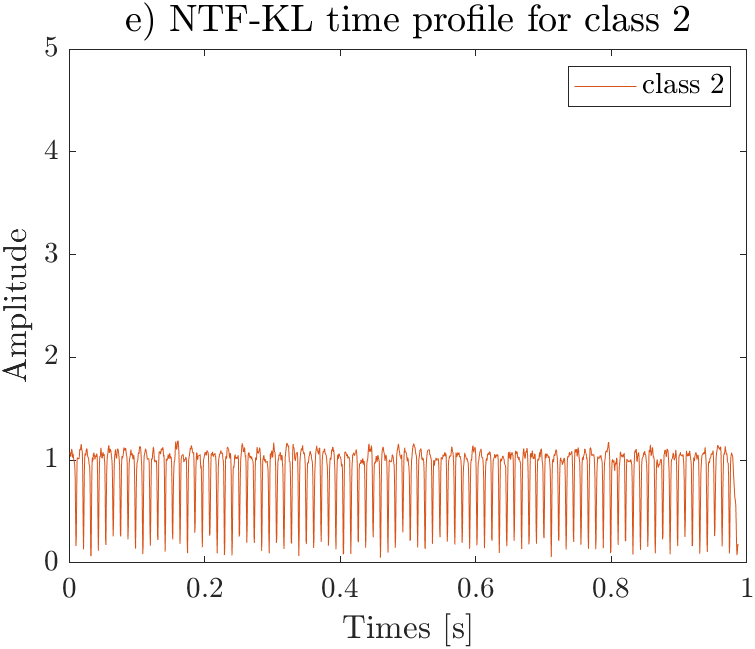}
    \end{subfigure}
    \hfill
    \begin{subfigure}[b]{.325\textwidth}
        \centering
        \includegraphics[width=\textwidth]{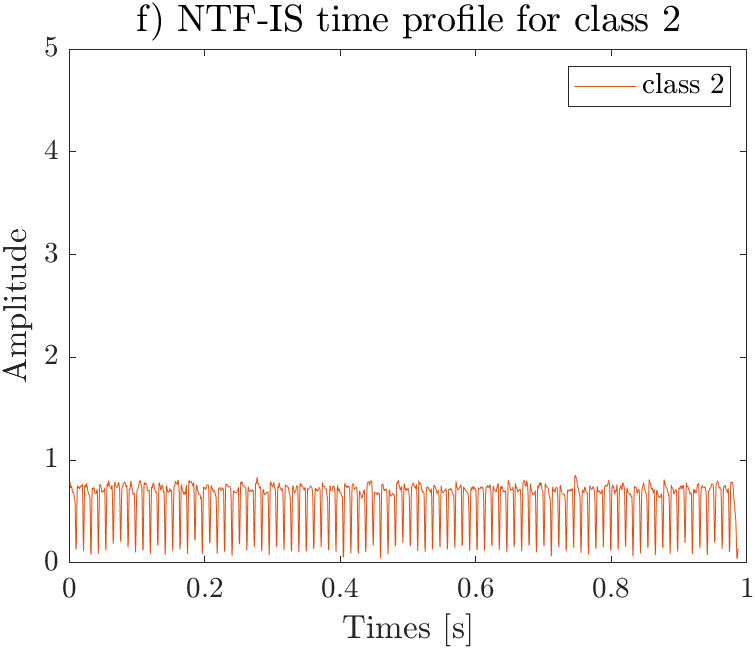}
    \end{subfigure}
    
    \caption{NTF time profiles of real vibration signal: a) NTF-EU class 1, b) NTF-KL class 1, c) NTF-IS class 1, d) NTF-EU class 2, e) NTF-KL class 2, f) NTF-IS class 2.}
    \label{fig:ntf_time_profiles}
\end{figure}

\subsection{NMF vs. NTF efficiency comparison}
\begin{figure}[H]
\centering

    \begin{subfigure}[b]{.325\textwidth}
        \centering
        \includegraphics[width=\textwidth]{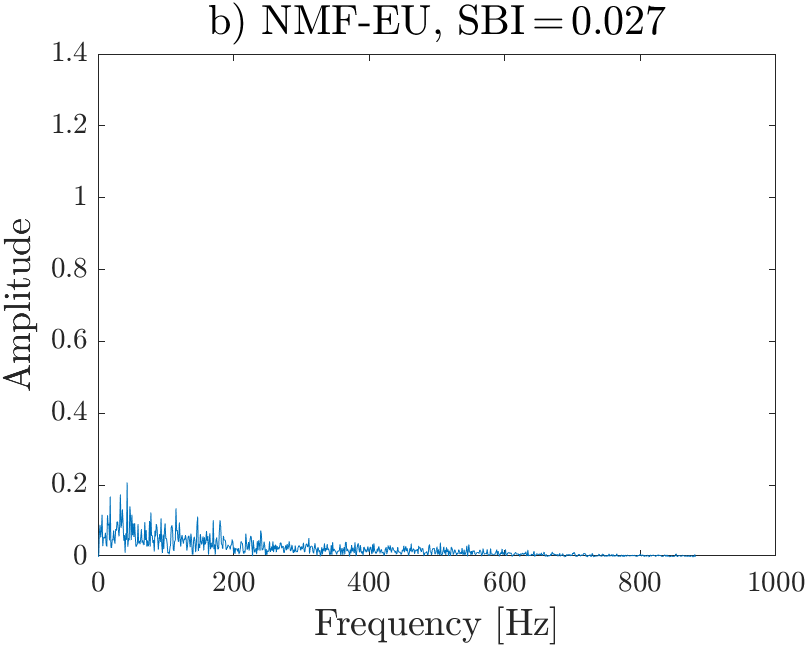}
    \end{subfigure}
    \hfill
    \begin{subfigure}[b]{.325\textwidth}
        \centering
        \includegraphics[width=\textwidth]{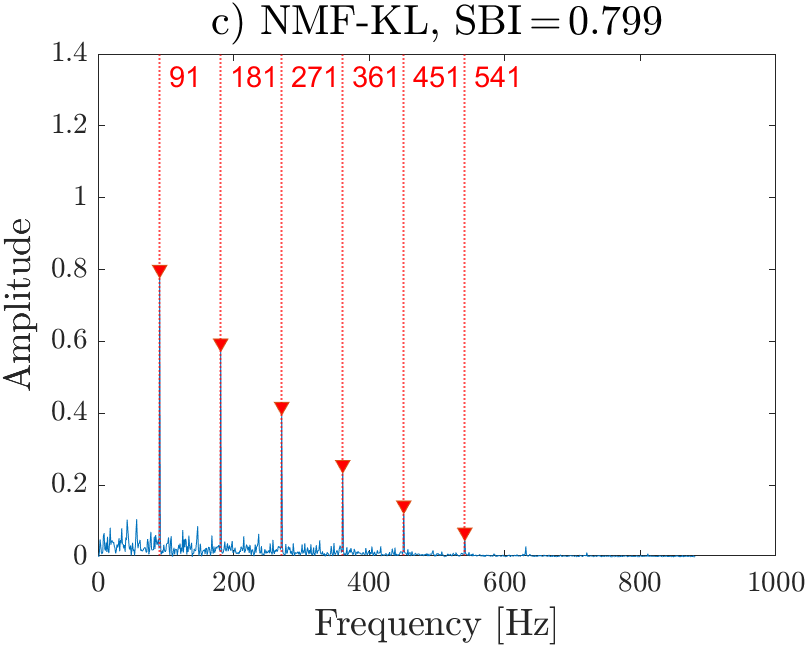}
    \end{subfigure}
    \hfill
    \begin{subfigure}[b]{.325\textwidth}
        \centering
        \includegraphics[width=\textwidth]{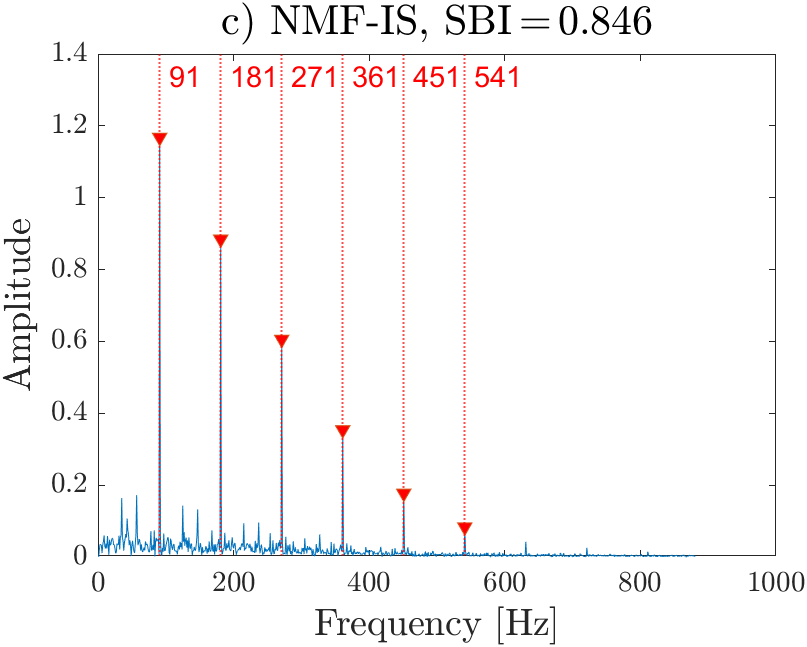}
    \end{subfigure}
    \newline
    
    \begin{subfigure}[b]{.325\textwidth}
        \centering
        \includegraphics[width=\textwidth]{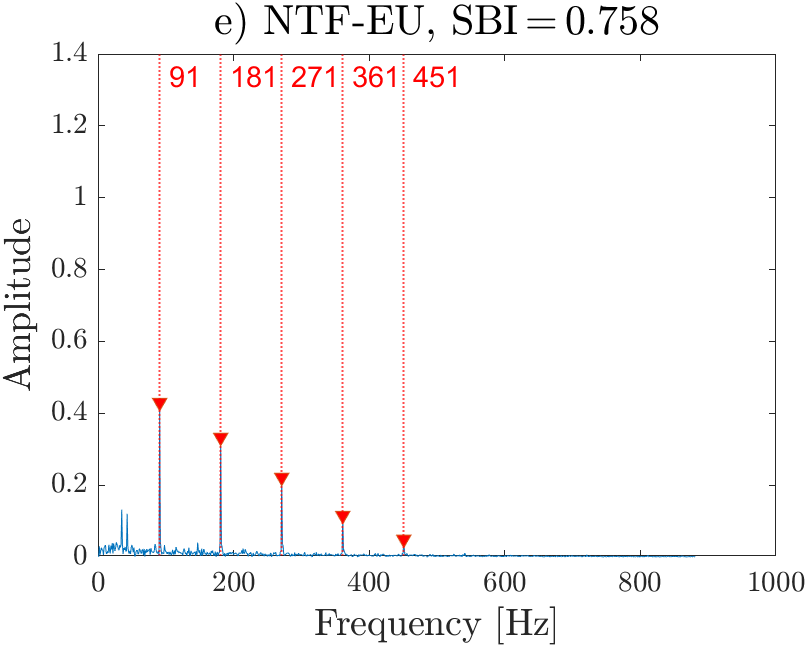}
    \end{subfigure}
    \hfill
    \begin{subfigure}[b]{.325\textwidth}
        \centering
        \includegraphics[width=\textwidth]{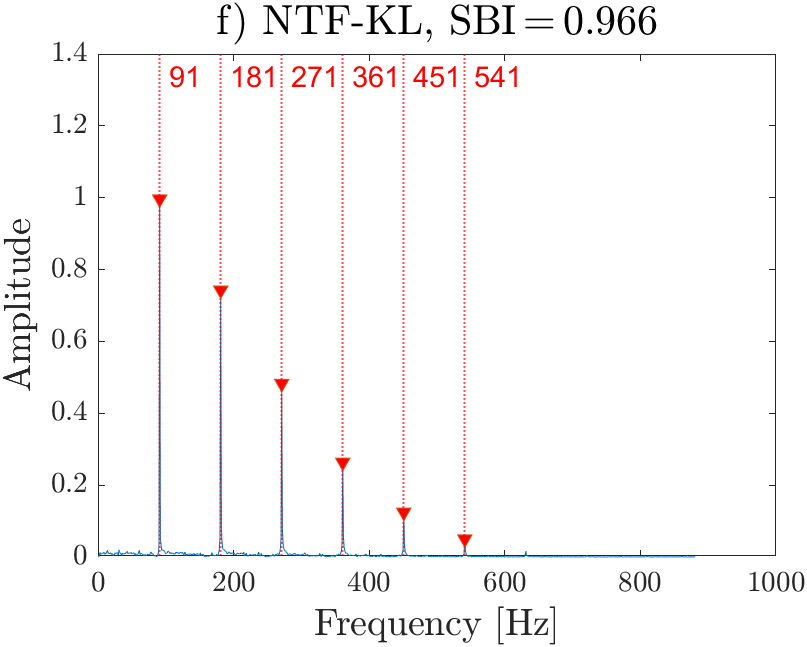}
    \end{subfigure}
    \hfill
    \begin{subfigure}[b]{.325\textwidth}
        \centering
        \includegraphics[width=\textwidth]{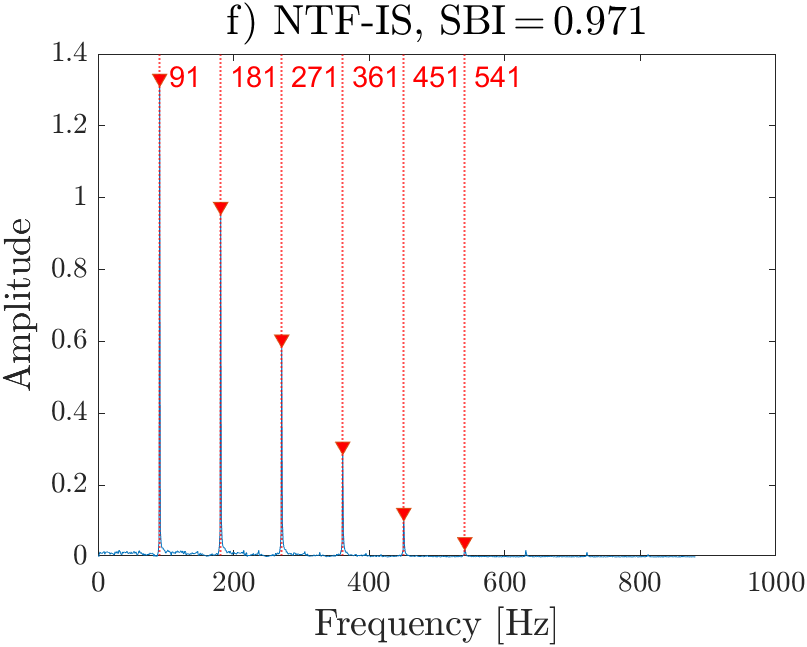}
    \end{subfigure}
    
    \caption{Spectra of NMF and NTF time profiles of real vibration signal: a) NMF-EU, b) NMF-KL, c) NMF-IS, d) NTF-EU, e) NTF-KL, f) NTF-IS.}
    \label{spectrums}
\end{figure}

Visual inspection of the results presented in Figures \ref{fig:nmf_ntf_real_freq_euclidean} -- \ref{fig:ntf_time_profiles} is informative and interesting. However, it is better to express the advantages of NTF quantitatively and to obtain the true confirmation that the extracted signal is periodic.
To compare the efficiency of NMF and NTF and highlight the advantages of NTF in the context of SOI extraction for the real signal, we can use the SBI measure again. The SBI is calculated according to formula (\ref{SBI_eq}). For the time profiles extracted from NMF and NTF, a spectrum is calculated, and then SBI is applied to measure how much “energy” of the SOI is presented in the extracted signal. The spectra calculated for time profiles from Figures \ref{fig:nmf_time_profiles} (a, b, c) and Figures \ref{fig:ntf_time_profiles} (a, b, c) are presented in Figures \ref{spectrums} (a-f). The detailed information on the algorithm and SBI is included in the title of each subplot. One may easily find that NTF is better than NMF (in the sense of the SBI value) for the same cost function. Moreover, the Itakura-Saito (IS) distance for both NMF and NTF provides the best results.

\section{Conclusions}
\label{sec:conclusions}
This study contributes to the area of vibration-based local damage detection in rolling element bearings. We proposed the NTF-based method that allows us to determine the informative frequency band, as well as to extract the latent cyclic impulsive component related to the damage. To the best of our knowledge, our study is the first work that fairly compares the NMF and NTF-based methods in the context of local damage detection. We presented and explained the advantages of using NTF in comparison to NMF. We also used the new cost functions (KL, IS) for both approaches, which significantly improved the quality of the results obtained. The theoretical analysis supporting the experiments provided the evidence that the update rules for NTF are more flexible in extracting the SOI from the mixture containing at least one non-Gaussian or non-stationary perturbing signal. The simulations with the synthetic signals confirmed these considerations, and showed that NTF offers significantly better efficiency than NMF in terms of the extraction of cyclic and impulsive signals at a very low level 
of SNR. The application of the discussed algorithms to real signals was also challenging, as a cyclic impulsive component was completely masked by noise and other interference. Thanks to NTF, the SOI has been extracted, and its pulse spectrum showed the strong peaks of appropriate fault frequencies and some weak background noise. The superiority of NTF over NMF was clearly seen for the Euclidean cost function (NMF failed, while NTF still provided good results). For the KL and IS cost functions, both NMF and NTF approaches provide the results of good quality but the background noise in the magnitude spectrum of the SOI is much weaker for the NTF-based method.

\section*{Acknowledgements}
This work is supported by National Center of Science under Sheng2 project No. UMO-2021/40/Q/ST8/00024 "NonGauMech - New methods of processing non-stationary signals (identification, segmentation, extraction, modeling) with non-Gaussian characteristics for the purpose of monitoring complex mechanical structures". 

\bibliography{bibliography}

\begin{thebibliography}{10}
\expandafter\ifx\csname url\endcsname\relax
  \def\url#1{\texttt{#1}}\fi
\expandafter\ifx\csname urlprefix\endcsname\relax\def\urlprefix{URL }\fi
\expandafter\ifx\csname href\endcsname\relax
  \def\href#1#2{#2} \def\path#1{#1}\fi

\bibitem{antoni2009cyclostationarity}
J.~Antoni, Cyclostationarity by examples, Mechanical Systems and Signal
  Processing 23~(4) (2009) 987--1036.

\bibitem{antoni2006spectral}
J.~Antoni, R.~Randall, The spectral kurtosis: application to the vibratory
  surveillance and diagnostics of rotating machines, Mechanical Systems and
  Signal Processing 20~(2) (2006) 308--331.

\bibitem{barszcz2009application}
T.~Barszcz, R.~B. Randall, Application of spectral kurtosis for detection of a
  tooth crack in the planetary gear of a wind turbine, Mechanical Systems and
  Signal Processing 23~(4) (2009) 1352--1365.

\bibitem{antoni2007fast}
J.~Antoni, Fast computation of the kurtogram for the detection of transient
  faults, Mechanical Systems and Signal Processing 21~(1) (2007) 108--124.

\bibitem{Xiang2015}
J.~Xiang, Y.~Zhong, H.~Gao, Rolling element bearing fault detection using
  {PPCA} and spectral kurtosis, Measurement 75 (2015) 180--191.

\bibitem{antoni2016info}
J.~Antoni, The infogram: Entropic evidence of the signature of repetitive
  transients, Mechanical Systems and Signal Processing 74 (2016) 73--94.

\bibitem{hebda2022infogram}
J.~Hebda-Sobkowicz, R.~Zimroz, A.~Wy{\l}oma{\'n}ska, J.~Antoni, Infogram
  performance analysis and its enhancement for bearings diagnostics in presence
  of non-gaussian noise, Mechanical Systems and Signal Processing 170 (2022)
  108764.

\bibitem{wang2016new}
T.~Wang, Q.~Han, F.~Chu, Z.~Feng, A new skrgram based demodulation technique
  for planet bearing fault detection, Journal of Sound and Vibration 385 (2016)
  330--349.

\bibitem{miao2017improvement_GINI}
Y.~Miao, M.~Zhao, J.~Lin, Improvement of kurtosis-guided-grams via gini index
  for bearing fault feature identification, Measurement Science and Technology
  28~(12) (2017) 125001.

\bibitem{schmidt2020methodology}
S.~Schmidt, A.~Mauricio, P.~S. Heyns, K.~C. Gryllias, A methodology for
  identifying information rich frequency bands for diagnostics of mechanical
  components-of-interest under time-varying operating conditions, Mechanical
  Systems and Signal Processing 142 (2020) 106739.

\bibitem{bozchalooi2007smoothness}
I.~S. Bozchalooi, M.~Liang, A smoothness index-guided approach to wavelet
  parameter selection in signal de-noising and fault detection, Journal of
  Sound and Vibration 308~(1-2) (2007) 246--267.

\bibitem{zhao2016detection}
M.~Zhao, J.~Lin, Y.~Miao, X.~Xu, Detection and recovery of fault impulses via
  improved harmonic product spectrum and its application in defect size
  estimation of train bearings, Measurement 91 (2016) 421--439.

\bibitem{mauricio2020improved}
A.~Mauricio, W.~A. Smith, R.~B. Randall, J.~Antoni, K.~Gryllias, Improved
  envelope spectrum via feature optimisation-gram (iesfogram): A novel tool for
  rolling element bearing diagnostics under non-stationary operating
  conditions, Mechanical Systems and Signal Processing 144 (2020) 106891.

\bibitem{hebda2020selection}
J.~Hebda-Sobkowicz, R.~Zimroz, A.~Wy{\l}oma{\'n}ska, Selection of the
  informative frequency band in a bearing fault diagnosis in the presence of
  non-gaussian noise—comparison of recently developed methods, Applied
  Sciences 10~(8) (2020) 2657.

\bibitem{antoni2004cyclostationary}
J.~Antoni, F.~Bonnardot, A.~Raad, M.~El~Badaoui, Cyclostationary modelling of
  rotating machine vibration signals, Mechanical systems and signal processing
  18~(6) (2004) 1285--1314.

\bibitem{kruczek2020detect}
P.~Kruczek, R.~Zimroz, A.~Wy{\l}oma{\'n}ska, How to detect the
  cyclostationarity in heavy-tailed distributed signals, Signal Processing
  (2020) 107514.

\bibitem{KRUCZEK2021107737}
P.~Kruczek, R.~Zimroz, J.~Antoni, A.~Wy{\l}oma{\'n}ska, Generalized spectral
  coherence for cyclostationary signals with alpha-stable distribution,
  Mechanical Systems and Signal Processing 159 (2021) 107737.

\bibitem{WODECKI2021108400}
J.~Wodecki, A.~Michalak, R.~Zimroz, Local damage detection based on vibration
  data analysis in the presence of gaussian and heavy-tailed impulsive noise,
  Measurement 169 (2021) 108400.

\bibitem{randall2011rolling}
R.~B. Randall, J.~Antoni, Rolling element bearing diagnostics - a tutorial,
  Mechanical Systems and Signal Processing 25~(2) (2011) 485--520.

\bibitem{lei2013review}
Y.~Lei, J.~Lin, Z.~He, M.~J. Zuo, A review on empirical mode decomposition in
  fault diagnosis of rotating machinery, Mechanical Systems and Signal
  Processing 35~(1-2) (2013) 108--126.

\bibitem{peng2004application}
Z.~Peng, F.~Chu, Application of the wavelet transform in machine condition
  monitoring and fault diagnostics: a review with bibliography, Mechanical
  systems and signal processing 18~(2) (2004) 199--221.

\bibitem{feng2013recent}
Z.~Feng, M.~Liang, F.~Chu, Recent advances in time--frequency analysis methods
  for machinery fault diagnosis: a review with application examples, Mechanical
  Systems and Signal Processing 38~(1) (2013) 165--205.

\bibitem{FabienM}
F.~Millioz, N.~Martin, Circularity of the {STFT} and spectral kurtosis for
  time-frequency segmentation in gaussian environment, {IEEE} Trans. Signal
  Process. 59~(2) (2011) 515--524.

\bibitem{wodecki2016combination}
J.~Wodecki, P.~Stefaniak, J.~Obuchowski, A.~Wylomanska, R.~Zimroz, Combination
  of principal component analysis and time-frequency representations of
  multichannel vibration data for gearbox fault detection., Journal of
  Vibroengineering 18~(4) (2016) 2167--2175.

\bibitem{lee1999learning}
D.~D. Lee, H.~S. Seung, Learning the parts of objects by non-negative matrix
  factorization, Nature 401~(6755) (1999) 788--791.

\bibitem{8610086}
Y.~Hao, L.~Song, M.~Wang, L.~Cui, H.~Wang, Underdetermined source separation of
  bearing faults based on optimized intrinsic characteristic-scale
  decomposition and local non-negative matrix factorization, IEEE Access 7
  (2019) 11427--11435.

\bibitem{9482220}
L.~Liang, X.~Ding, H.~Wen, L.~Shan, Periodic impulse feature separation by
  combination of bi-frequency map and non-negative matrix factorization, in:
  2021 IEEE 4th Advanced Information Management, Communicates, Electronic and
  Automation Control Conference (IMCEC), Vol.~4, 2021, pp. 1360--1364.
\newblock \href {https://doi.org/10.1109/IMCEC51613.2021.9482220}
  {\path{doi:10.1109/IMCEC51613.2021.9482220}}.

\bibitem{8942862}
H.~Luo, L.~Song, M.~Wang, H.~Wang, L.~Cui, Compound faults diagnosis method
  based on adaptive gst-nmf for rolling bearing, in: 2019 Prognostics and
  System Health Management Conference (PHM-Qingdao), 2019, pp. 1--6.

\bibitem{Fu2019NonnegativeMF}
X.~Fu, K.~Huang, N.~D. Sidiropoulos, W.-K. Ma, Nonnegative matrix factorization
  for signal and data analytics: Identifiability, algorithms, and applications,
  IEEE Signal Processing Magazine 36 (2019) 59--80.

\bibitem{Casalino2016}
G.~Casalino, N.~Del~Buono, C.~Mencar, Nonnegative Matrix Factorizations for
  Intelligent Data Analysis, Springer Berlin Heidelberg, Berlin, Heidelberg,
  2016, pp. 49--74.

\bibitem{cichocki2009nonnegative}
A.~Cichocki, R.~Zdunek, A.~H. Phan, S.-i. Amari, Nonnegative matrix and tensor
  factorizations: applications to exploratory multi-way data analysis and blind
  source separation, John Wiley \& Sons, 2009.

\bibitem{Wang_2013}
Y.-X. Wang, Y.-J. Zhang, Nonnegative matrix factorization: A comprehensive
  review, IEEE Trans. on Knowl. and Data Eng. 25~(6) (2013) 1336--1353.

\bibitem{gillis2020nonnegative}
N.~Gillis, Nonnegative Matrix Factorization, Data Science, Society for
  Industrial and Applied Mathematics (SIAM), 2020.

\bibitem{wodecki2019novel}
J.~Wodecki, P.~Kruczek, A.~Bartkowiak, R.~Zimroz, A.~Wy{\l}oma{\'n}ska, Novel
  method of informative frequency band selection for vibration signal using
  nonnegative matrix factorization of spectrogram matrix, Mechanical Systems
  and Signal Processing 130 (2019) 585--596.

\bibitem{Wodecki2020}
J.~Wodecki, A.~Michalak, R.~Zimroz, A.~Wy{\l}oma{\'n}ska, Separation of
  multiple local-damage-related components from vibration data using
  nonnegative matrix factorization and multichannel data fusion, Mechanical
  Systems and Signal Processing 145 (2020) 106954.

\bibitem{wodecki2019impulsive}
J.~Wodecki, A.~Michalak, R.~Zimroz, T.~Barszcz, A.~Wy{\l}oma{\'n}ska, Impulsive
  source separation using combination of nonnegative matrix factorization of
  bi-frequency map, spatial denoising and monte carlo simulation, Mechanical
  Systems and Signal Processing 127 (2019) 89--101.

\bibitem{griffin1984signal}
D.~Griffin, J.~Lim, Signal estimation from modified short-time {F}ourier
  transform, IEEE Transactions on Acoustics, Speech, and Signal Processing
  32~(2) (1984) 236--243.

\bibitem{Shashua2005}
A.~Shashua, T.~Hazan, Non-negative tensor factorization with applications to
  statistics and computer vision, in: Proc. of the 22-th International
  Conference on Machine Learning, Bonn, Germany, 2005.

\bibitem{carroll1989fitting}
J.~D. Carroll, G.~De~Soete, S.~Pruzansky, Fitting of the latent class model via
  iteratively reweighted least squares candecomp with nonnegativity
  constraints, in: Multiway data analysis, 1989, pp. 463--472.

\bibitem{FitzGerald2008}
D.~FitzGerald, M.~Cranitch, E.~Coyle, Extended nonnegative tensor factorisation
  models for musical sound source separation, Comput Intell Neurosci 872425
  (2008) 1--15.

\bibitem{6588559}
M.~Figueiredo, B.~Ribeiro, A.~de~Almeida, Electrical signal source separation
  via nonnegative tensor factorization using on site measurements in a smart
  home, IEEE Transactions on Instrumentation and Measurement 63~(2) (2014)
  364--373.

\bibitem{8497054}
F.~Xiong, Y.~Qian, J.~Zhou, Y.~Y. Tang, Hyperspectral unmixing via total
  variation regularized nonnegative tensor factorization, IEEE Transactions on
  Geoscience and Remote Sensing 57~(4) (2019) 2341--2357.

\bibitem{app9183642}
L.~Liang, H.~Wen, F.~Liu, G.~Li, M.~Li, Feature extraction of impulse faults
  for vibration signals based on sparse non-negative tensor factorization,
  Applied Sciences 9~(18) (2019).

\bibitem{Kolda08}
T.~G. Kolda, B.~W. Bader, Tensor decompositions and applications, SIAM Review
  51~(3) (2009) 455--500.

\bibitem{Carroll_Chang_70}
J.~D. Carroll, J.~J. Chang, Analysis of individual differences in
  multidimensional scaling via an n-way generalization of {E}ckart-{Y}oung
  decomposition, Psychometrika 35 (1970) 283--319.

\bibitem{Harshman1970}
R.~A. Harshman, Foundations of the {PARAFAC} procedure: Models and conditions
  for an \textquotedblleft explanatory\textquotedblright multimodal factor
  analysis, UCLA Working Papers in Phonetics 16 (1970) 1--84.

\end{thebibliography}

\end{document}